\documentclass{ar-1col}

\usepackage{latexsym}
\usepackage{amssymb}
\usepackage{epsfig,amsmath,graphics}
\usepackage{rotating}%
\usepackage{url}
\usepackage[latin1]{inputenc}

\newcommand{\lsim}{\mathrel{\mathop{\kern 0pt \rlap
  {\raise.2ex\hbox{$<$}}}
  \lower.9ex\hbox{\kern-.190em $\sim$}}}
\newcommand{\gsim}{\mathrel{\mathop{\kern 0pt \rlap
  {\raise.2ex\hbox{$>$}}}
  \lower.9ex\hbox{\kern-.190em $\sim$}}}

\newcommand{\alt}{\mathrel{\mathop{\kern 0pt \rlap
  {\raise.2ex\hbox{$<$}}}
  \lower.9ex\hbox{\kern-.190em $\sim$}}}
\newcommand{\agt}{\mathrel{\mathop{\kern 0pt \rlap
  {\raise.2ex\hbox{$>$}}}
  \lower.9ex\hbox{\kern-.190em $\sim$}}}

 \newcommand{\I}{\rm{i}}

\newcommand{\gagamma}{g_{a \gamma \gamma}}
\newcommand{\ckcs}{\text{counts}~keV$^{-1}$~cm$^{-2}$~s$^{-1}$}

\jname{Xxxx. Xxx. Xxx. Xxx.}
\jvol{AA}
\jyear{YYYY}
\doi{10.1146/((please add article doi))}

\begin{document}

% Page header
\markboth{Peter W. Graham, Igor G. Irastorza, Steven K. Lamoreaux, Axel Lindner, and Karl A. van Bibber}{Experimental Axion Searches}

% Title
\title{Experimental Searches for the Axion and Axion-like Particles}

%Authors, affiliations address.
\author{Peter W. Graham,$^1$ Igor G. Irastorza,$^2$ Steven K. Lamoreaux,$^3$ Axel Lindner,$^4$ and Karl A. van Bibber,$^5$
\affil{$^1$Corresponding author, Stanford Institute for Theoretical Physics, Stanford University, Stanford, CA, USA, 94305}
\affil{$^2$Laboratorio de Física Nuclear y Astropartículas, Departamento de Física Teórica, Universidad de Zaragoza, 50009 Zaragoza, Spain}
\affil{$^3$Department of Physics, Yale University, New Haven, CT, USA, 06511}
\affil{$^4$DESY, Notkestra\ss e 85, 22607 Hamburg, Germany}
\affil{$^5$Department of Nuclear Engineering, University of California Berkeley, Berkeley CA, USA 94720}}

%Abstract
\begin{abstract}
Four decades after its prediction, the axion remains the most compelling solution to the Strong-CP problem and a well-motivated dark matter candidate, inspiring a host of elegant and ultrasensitive experiments based on axion-photon mixing.  This report reviews the experimental situation on several fronts. The microwave cavity experiment is making excellent progress in the search for dark matter axions in the microelectronvolt range and may be plausibly extended up to 100 $\mu$eV.  Within the past several years however, it has been realized that axions are pervasive throughout string theories, but with masses that fall naturally in the nanoelectronvolt range, for which a NMR-based search is under development.  Searches for axions emitted from the Sun's burning core, and purely laboratory experiments based on photon regeneration have both made great strides in recent years, with ambitious projects proposed for the coming decade.  Each of these campaigns has pushed the state of the art in technology, enabling large gains in sensitivity and mass reach.  Furthermore each modality has also been exploited to search for more generalized axion-like particles, that will also be discussed in this report.  We are hopeful, even optimistic, that the next review of the subject will concern the discovery of the axion, its properties, and its exploitation as a probe of early universe cosmology and structure formation. 
\end{abstract}

%Keywords, etc.
\maketitle

%Table of Contents
\tableofcontents

% Heading 1
\section{Introduction}

Since its prediction in 1978 \cite{Weinberg:1977ma, Wilczek:1977pj}, there has been a steady crescendo of interest in the axion, and a coordinated global effort to find it is now finally taking shape.  Should it ultimately be discovered, it would finally resolve one of the last outstanding questions in the Standard Model of particle physics, namely it would validate the Peccei-Quinn mechanism to protect the strong interaction from CP-violating effects \cite{Peccei:1977hh, Peccei:1977ur}, as evidenced by the absence of a neutron electric dipole moment.   Furthermore, it may be discovered as the dark matter halo of our Milky Way galaxy, thus answering the question of what constitutes the predominant form of matter in our Universe.
This will be an experimental review, focusing on new concepts and developments since the last such report in this journal \cite{Asztalos:2006kz} and in others \cite{Rosenberg:2000wb, Bradley:2003kg}.  The reader is referred to several excellent theory reviews \cite{Kim:1986ax, Cheng:1987gp, Turner:1989vc, Raffelt:1990yz, Kim:2008hd} that are still largely up to date. There is one new development of interest to our present purpose however, namely the realization that any string theory contains several and perhaps a large number of axion-like particles, although they are extremely light, of order neV.  This will be discussed briefly in Sec. \ref{section: CASPEr}, before describing the NMR-based experiment to search that mass region.
Nevertheless, some theoretical preliminaries are in order, therefore this review will begin with the basic physics underlying the axion and its phenomenology in Sec. \ref{sec: strong cp and axion}.  Constraints on the axion's mass and couplings primarily from its cosmological production and astrophysics will be introduced at the beginning of Sec.~\ref{sec: dm axion searches} to prepare for the discussion of the microwave cavity experiment.  Axion-photon mixing within a magnetic field in the relativistic limit relevant both to the solar searches and laboratory experiments will be presented at the beginning of Sec.~\ref{sec: solar axions}, along with axion emission from the Sun's burning core. 

While this review primarily concerns the axion solving the strong-CP problem, more generalized axion-like particles (ALPs) accessible to these experiments will also be dealt with.  As we are writing from the experimental perspective, after all, one should be open to surprises!

%Please begin the main text of your article here. This is a template file for AR Journals.

% Text Box
%\begin{textbox}
%\section{TEXT BOX HEAD}
%Text box text. Text box text. Text box text. Text box text. Text box text. Text box text.
%\subsection{Text Box Sub-head}
%Text box text. Text box text. Text box text. Text box text. Text box text. Text box text.
%\subsubsection{Text Box Subsub-head}
%Text box text. Text box text. Text box text. Text box text. Text box text. Text box text.
%\end{textbox}

%\begin{textbox}
%Text box text. Text box text. Text box text. Text box text. Text box text. Text box text.
%Text box text. Text box text. Text box text. Text box text. Text box text. Text box text.
%Text box text. Text box text. Text box text. Text box text. Text box text. Text box text.
%\end{textbox}

%Heading 1
%\section{FIRST-ORDER HEADING}
%This is dummy text. This is dummy text. This is dummy text. This is dummy text. This is dummy text. This is dummy text. This is dummy text. This is dummy text. This is dummy text. This is dummy text. This is dummy text. This is dummy text.

% Heading 2
%\subsection{Second-Order Heading}
%This is dummy text. This is dummy text. This is dummy text. This is dummy text.

% Heading 3
%\subsubsection{Third-Order Heading}
%This is dummy text. This is dummy text. This is dummy text. This is dummy text. 

% Heading 4
%\paragraph{Fourth-Order Heading:} Fourth-order headings are placed as part of the paragraph.

\section{Strong-CP and the Axion}
\label{sec: strong cp and axion}

The gauge sector of the Standard Model of electroweak interactions is among the
most successful theories in the history of physics, while the flavor sector is
still incomplete with known issues and uncertainties, {\it e.g.}, the neutrino mass spectrum.  Perhaps the most 
mysterious issue is that of the
so-called strong $CP$ problem (combined charge conjugation $C$ and parity inversion $P$ symmetry, or equivalently time reversal $T$ symmetry) -- why
does the quantum chromodynamic (QCD) Lagrangian conserve $CP$ symmetry (or equivalently time reversal $T$ symmetry)
apparently perfectly, to within extraordinarily strict experimental limits, when there is no fundamental reason to 
exclude possible symmetry nonconserving interactions?  

As described in \cite{skl0}, 
this problem can be understood as follows:
The Lagrangian of the electromagnetic field
\begin{equation}
-\ {1\over 4}F_{{\mu}\nu}F_{{\mu}\nu} = {1\over 2} (\vec E^2 - \vec B^2)
\end{equation}
%\[-\ {1\over 4}F_{{\mu}\nu}F_{{\mu}\nu} = {1\over 2} (\vec E^2 - \vec B^2)\]
can comprise other Lorentz scalars, such as \cite{skl1}
\begin{equation}
F_{{\mu}\nu}\tilde{F}_{{\mu}\nu},\;\;\;\;\tilde{F}_{{\mu}\nu} =
{1\over 2}\epsilon_{{\mu}\nu\kappa\lambda}F_{\kappa\lambda}.
\end{equation}
%\[F_{{\mu}\nu}\tilde{F}_{{\mu}\nu},\;\;\;\;\tilde{F}_{{\mu}\nu} =
%{1\over 2}\epsilon_{{\mu}\nu\kappa\lambda}F_{\kappa\lambda}.\]
This scalar violates both $P$ and $T$ invariance, as evident from its
three dimensional form:
\begin{equation}
F_{{\mu}\nu}\tilde{F}_{{\mu}\nu} = - 4 \vec E \cdot \vec B.
\end{equation}
%\[F_{{\mu}\nu}\tilde{F}_{{\mu}\nu} = - 4 \vec E \cdot \vec B.\]
However this scalar generates no
observable effects in electrodynamics because it is a 4-divergence and the fields fall
off rapidly toward infinity.

The corresponding possible $P$
and $T$ violating term in the QCD Lagrangian is usually written as
\begin{equation}\label{th}
L_{\theta} =\, -\, \theta\,(\alpha_s/8{\pi})\,\tilde{G}^a_{{\mu}\nu} G^a_{{\mu}\nu}
\end{equation}
and is called the $\theta$ term, where $\alpha_s\sim 1$ is the
coupling constant for the gluon field $G$, and is the QCD analogue of the fine structure constant
$\alpha = 1/137$ in electrodynamics. While the four-divergence of this term can remain
zero, implying that dynamics are not directly affected by it, both the chiral anomaly and
the self-interaction of the gluon vector potential
field configurations, which do not fall off rapidly enough at
infinity, can lead to observable effects.  

Most famously, the $\theta$ term provides a contribution to the neutron electric dipole moment (EDM) which is estimated in
\cite{skl2,skl3}. First, a chiral rotation $\psi
\rightarrow \exp{(-\I\gamma_5 \theta)} \psi$ of the quark spinor fields
$\psi$ transforms the $\theta$ term away. Under this rotation, the
mass term in the Hamiltonian of the light quarks, $u$, $d$, $s$,
\begin{equation}
 m_u \bar u u + m_d \bar d d + m_s \bar s s 
\end{equation}
%\[ m_u \bar u u + m_d \bar d d + m_s \bar s s \]
acquires a $CP$-odd term
\begin{equation}\label{hcp}
\delta H_{CP} =
i \theta\, {m_u m_d m_s\over m_u m_d + m_u m_s + m_d m_s}\,
(\bar u \gamma_5 u + \bar d \gamma_5 d + \bar s \gamma_5 s).
\end{equation}
Because the $s$ quark is much heavier
than $u$ and $d$, the mass factor simplifies to
\begin{equation}
{m_u m_d m_s\over m_u m_d + m_u m_s + m_d m_s} \approx
{m_u m_d\over m_u + m_d}.
\end{equation}
%\[{m_u m_d m_s\over m_u m_d + m_u m_s + m_d m_s} \approx
%{m_u m_d\over m_u + m_d}.\]
The $CP$-odd ${\pi} NN$ vertex generated by this Hamiltonian
can be transformed by use of the PCAC technique to
\begin{equation}
\langle{\pi}^a N_{\rm f}\,|\delta H_{CP}|\,N_{\rm i}\rangle = \,
-\, \theta\, \frac{m_u m_d}{m_u + m_d}\, \frac{\sqrt 2}{f_{{\pi}}} \,
\langle N_{\rm f}\,|\,\bar q \tau^a q\,|\,N_{\rm i}\rangle
\end{equation}
where $\tau^a$ is the isotopic spin operator and $f_{{\pi}}=130$ MeV is
the pion decay constant. The nucleon matrix element $\langle p\,|\,\bar u
d\,|\,n\rangle$ is related through
SU$(3)$ symmetry to the mass splitting in the baryon octet:
\begin{equation}\label{su3}
\langle p\,|\,\bar u d\,|\,n \rangle\,
=\,\bar p n \,\frac{M_{\Xi} - M_{\Sigma}}{m_s}\,\approx \,\bar p n
\end{equation}
where $p$ and $n$ are the Dirac spinors of the proton and neutron;
$M_{\Xi}$ and $M_{\Sigma}$ are the masses of the $\Xi$  and
$\Sigma$ hyperons, respectively. The full ${\pi} NN$ interaction can be
now be described by an effective Hamiltonian
\begin{equation}
H_{{\pi} NN}=\,\vec{{\pi}}\,\bar N \,\vec{\tau}\,(\,i \,\gamma_5
g_{{\pi} NN}\,+\,\bar{g}_{{\pi} NN})\,N,
\end{equation}
where the $CP$-odd constant is
\begin{equation}
\bar{g}_{{\pi} NN}\,
=\,- \theta\, {m_u m_d\over m_u + m_d}\,{\sqrt 2\over f_{{\pi}}}\,
{M_{\Xi} - M_{\Sigma}\over m_s}\,\approx  -\,0.027 \theta.
\end{equation}
The $CP$-even ${\pi} NN$ constant in the effective
Hamiltonian is known,
\begin{equation}
g_{{\pi} NN}\,=\,13.6.
\end{equation}

In \cite{skl3}, a crucial observation is made. In the chiral
limit $m_{{\pi}}\rightarrow 0$ the neutron EDM can be expressed
exactly via $\bar{g}_{{\pi} NN}$ and $g_{{\pi} NN}$. In this limit,
there are only two diagrams that are
singular in the pion mass and thus contribute to the EDM. In these diagrams, a ${\pi} NN$ vertex is the strong
pseudoscalar coupling, with the coupling constant $g_{{\pi} NN}\sqrt 2$, and
the second is a $CP$-odd scalar, with the coupling constant
$\bar{g}_{{\pi} NN}\sqrt 2$. The contribution of these diagrams to the
neutron EDM is
\begin{equation}\label{chir}
d_{\rm n}\,=\,{|e|\over m_{\rm p}}\,{g_{{\pi} NN}\, \bar{g}_{{\pi} NN}\over
4{\pi}^2}\,
\ln{\frac{m_{\rho}}{m_{{\pi}}}}=-3.3\times 10^{-16}\theta\ [e \cdot {\rm cm}].
\end{equation}
The choice of the $\rho$ meson mass $m_{\rho}=\,770$ MeV as the
typical hadronic scale at which the logarithmic integral is cut off is somewhat
arbitrary. The chiral parameter,  the logarithm in Eq. (\ref{chir}),
is not large for any reasonable cut-off and is only 1.7 when
$m_{\rho}$ is used.  Due to the absence of other
terms logarithmic in $m_{{\pi}}$, a coincidental mutual cancellation between
this contribution and possible others appears as unlikely. Therefore
Eq. (\ref{chir}) can be considered a conservative
estimate of the neutron EDM.

Combining this estimate with the most sensitive experimental result given in Ref. \cite{skl4},
$d_{\rm n}< 2.9\times 10^{-26}\ e \cdot $cm,
sets a very strict upper limit for the $CP$-odd QCD parameter
\begin{equation}\label{thex}
|\theta|\,<\,9\times 10^{-11}.
\end{equation}

This is the strong $CP$ problem: There is no natural
explanation for the extreme smallness of the parameter $\theta$ and
is considered to be a fine-tuning problem.
Indeed, it appears as particularly unnatural when one considers
that $\theta$ is renormalized by other $CP$-odd
interactions (e.g., known $CP$ violation in $K$ meson decay), 
and in general its renormalization can be infinite. It has been shown that, in
the Standard Model, the induced contributions to
$\theta$ almost certainly diverge logarithmically starting at 14$^\text{th}$(!) order in
the electroweak coupling constant \cite{skl5,skl6}. Therefore, as a technicality, 
$\theta$ cannot be calculated.

One solution to this problem, which is quite
obvious from Eq. (\ref{hcp}), is to assume that one of the quark
masses is zero or very small (lighter than a neutrino). This assumption apparently 
contradicts experimental data, although that is not entirely certain.  

A solution to the strong $CP$ problem that has wide interest, particularly in the context of this Review,
is to leave the $\theta$ term as it is, but to somehow make it irrelevant.
This is achieved by introducing an extra global symmetry into the
theory \cite{Peccei:1977hh,Peccei:1977ur}, {\it i.e.}, by considering $\theta$ as a field, not a fixed parameter. 
Such a symmetry leads in turn to the prediction of a new light pseudoscalar particle, the axion
\cite{Weinberg:1977ma,Wilczek:1977pj}.  In a restricted sense, the axion plays a role similar to the Higgs boson;
the $CP$ violating interaction cannot be calculated in the context of the Standard Model and in
fact diverges without the axion, while electroweak interactions beyond first order diverge without the (now discovered!) Higgs boson.  Of course
it would be incorrect to say that the axion is as well motivated as the Higgs boson, however the
parallelism should be noted. 

The mass of the axion is unknown, but is bound from both above and from
below by experiments and observations.

%%%%%%%%%%%%%%%%%%% End of Steve's part, beginning of Karl's %%%%%%%%%%%%%%

Essentially all of the physics of the axion depends on a large unknown energy scale $f_a$, at which Peccei-Quinn symmetry is broken; in the low-energy limit of their theory, the non-perturbative vacuum structure of QCD drives the parameter of the CP-violating term, to the CP-conserving minimum, with the axion resulting from the remnant oscillations of the axion field about this minimum.  The mass of the axion is given by \cite{Cheng:1987gp, Turner:1989vc, Raffelt:1990yz}:
\begin{equation}
m_a \approx 6 \, \text{eV} \left( \frac{10^6 \, \text{GeV}}{f_a} \right)
\end{equation}
In its original form, the PQ symmetry-breaking scale was posited (for no compelling reason) to be of order the electroweak scale, $f_\text{EW}$, implying very heavy axions ($\sim$ 100 keV) which were quickly ruled out by accelerator- and reactor-based experiments.  More general models were then constructed of much higher values of $f_a$ and much smaller $m_a$.

Generically, all the couplings of the axion to radiation and matter are also inversely proportional to $f_a$.  The axion-photon coupling is of special interest here, as virtually all of the most sensitive search strategies are based on the coherent mixing of axions and photons in a strong magnetic field. Being a pseudoscalar ($J^\pi = 0^-$), the axion has a two-photon coupling whose strength is given by:
\begin{equation}
g_{a\gamma\gamma} = \frac{\alpha g_\gamma}{\pi f_a}
\end{equation}
where $g_\gamma$ is a dimensionless model-dependent parameter of order unity; $g_\gamma = -0.97$ in the KSVZ model \cite{Kim:1979if, Shifman:1979if}, and $g_\gamma = 0.36$ in the DFSZ model \cite{Dine:1981rt, Zhitnitsky:1980tq}. These values are representative within broad classes of hadronic and GUT-inspired axions, underlining an important feature of the theory that the dimensionless axion-photon coupling $g_\gamma$ is highly insensitive to the specific axion model. This observation extends even to axions ubiquitous within string theory, and which are perhaps even intrinsic to their structure \cite{Svrcek:2006yi}\footnote{This paper erroneously implies $g_\gamma^\text{STRING} = (1/4) g_\gamma^\text{DFSZ}$; however the author subsequently clarified that $g_\gamma$ within string models should be {\it exactly} equal to that of the most generic model, DFSZ, and represents a strict lower limit to the coupling.  E. Witten, private communication (2007)}.

%  Historically, the conversion of a single photon to a pseudoscalar in an external electromagnetic field (representing a sea of virtual photons) was known as the Primakoff effect, by which the first accurate determinations of the $\pi^0$ lifetime were made.

%\section{Axion-Photon Mixing}
%\label{sec: axion photon}

\subsection{Axion-like particles and other WISPs}
\label{sc:wisps}
As explained above there is a strong physics case for the axion, but it may be just a first representative of a new particle family of so called Weakly Interacting Slim Particles (WISPs). Such WISPs are motivated for example by string-theory inspired extensions of the Standard Model, which predict, among others, the existence of axion-like particles (ALPs) and hidden photons (HPs)~\cite{Jaeckel:2010ni}.
%ALPs do not originate from a Peccei-Quinn symmetry (by definition), so they are not related to the CP problem of QCD and their masses and coupling strengths are not determined by the same energy scales.
While entirely unrelated to the Strong-CP problem, ALPs and HPs may also be viable dark matter candidates~\cite{ Arias:2012az, Ringwald:2012hr}. ALPs are of special interest, because there are a number of different astrophysics phenomena like the transparency of the universe to TeV photons~\cite{ Meyer:2013pny, Rubtsov:2014uga} and evolution of stars~\cite{ Ayala:2014pea, Bertolami:2014wua}, which might point to the existence of very light ALPs with a coupling strength detectable at the next generation of helioscopes and purely laboratory experiments. It should be stressed that the ongoing discussion on the validity of Naturalness (questioning arguments for the existence of new particles based on fine tuning issues) triggered by the results of the first run of LHC (see~\cite{Altarelli:2014ola,Feng:2013pwa} for example) does not influence the hints for WISPs as sketched here.
Thus the parameter space for very light and very feebly interacting particles is opening up next to the axion region and experiments are encouraged to widen their field of view accordingly.

\section{Searches for Dark Matter Axions}
\label{sec: dm axion searches}

\subsection{Axionic dark matter}

A sufficiently light axion represents an excellent dark-matter candidate \cite{Sikivie:2006ni}, as its density relative to the critical density of the universe is given by
\begin{equation}
\Omega_a \approx \left( \frac{6 \, \mu \text{eV}}{m_a} \right)^\frac{7}{6}
\end{equation}
An axion of $m_a \approx 20 \, \mu \text{eV}$ (within a factor of $\sim 2$) would thus account for the entire dark matter density of the universe, $\Omega_m \approx 0.23$.  Without tuning of the initial misalignment angle much lighter axions would overclose the universe, and therefore $m_a \approx 1 \, \mu \text{eV}$ may be taken as a strong lower limit on the axion mass, $m_a$.  There has been a lingering controversy about the relative contribution between axion production from the vacuum realignment mechanism and radiation from topological defects (axion strings, domain walls).  A definitive resolution is unlikely soon, but the current general consensus is that the sum of all contributions to $\Omega_a$ could push the axion mass corresponding to $\Omega_\text{DM}$ two orders of magnitude higher.  As we are presently interested only in establishing conservative ranges for experiments searches should plan on reaching the $\mu$eV scale.  Cosmology also provides an upper bound on the axion mass by the production of thermal axions, the hot dark matter limit of about 1 eV, but this does not concern us here.

Stellar evolution also places strict limits on axions \cite{Raffelt:2006cw}.
%\cite{Raffelt:1996wa, Raffelt:2006cw}
  Axions of mass exceeding $\sim 16$ meV would have quenched the neutrino pulse observed from SN1987a, thus bounding the axion mass from above.  The best indirect stellar bound on the axion-photon coupling comes from 
Galactic Globular Clusters, $g_{a \gamma\gamma} < 0.66 \times 10^{-10} \, \text{GeV}^{-1}$, recently surpassing the best direct stellar bound,
%Horizontal Branch Stars, $g_{\gamma\gamma} \approx 10^{-10} \, \text{GeV}^{-1}$, now surpassed by the best direct stellar bound,
the CAST search for solar axions, $g_{a \gamma \gamma} \approx 0.88 \times 10^{-10} \, \text{GeV}^{-1}$ discussed in Sec. \ref{sec: solar axions}.  For Peccei-Quinn axions, such couplings correspond to masses far in excess of the open mass region, i.e. $10^{-6} \, \text{eV} < m_a < 10^{-2} \, \text{eV}$, and thus are not germane to the dark matter problem.  There has been a modest literature in recent years suggesting that the luminosity function of white dwarfs (degenerate stars undergoing gravothermal cooling) may require an additional cooling mechanism that could be accounted for by axions in the $\sim 10$ meV range \cite{Isern:2008nt, Isern:2010wz, Corsico:2012sh}. The evidence is far from compelling, not to mention that such masses are beginning to encroach in the region disfavored by SN1987a.

\begin{figure}
\begin{center}
\includegraphics[width= \columnwidth]{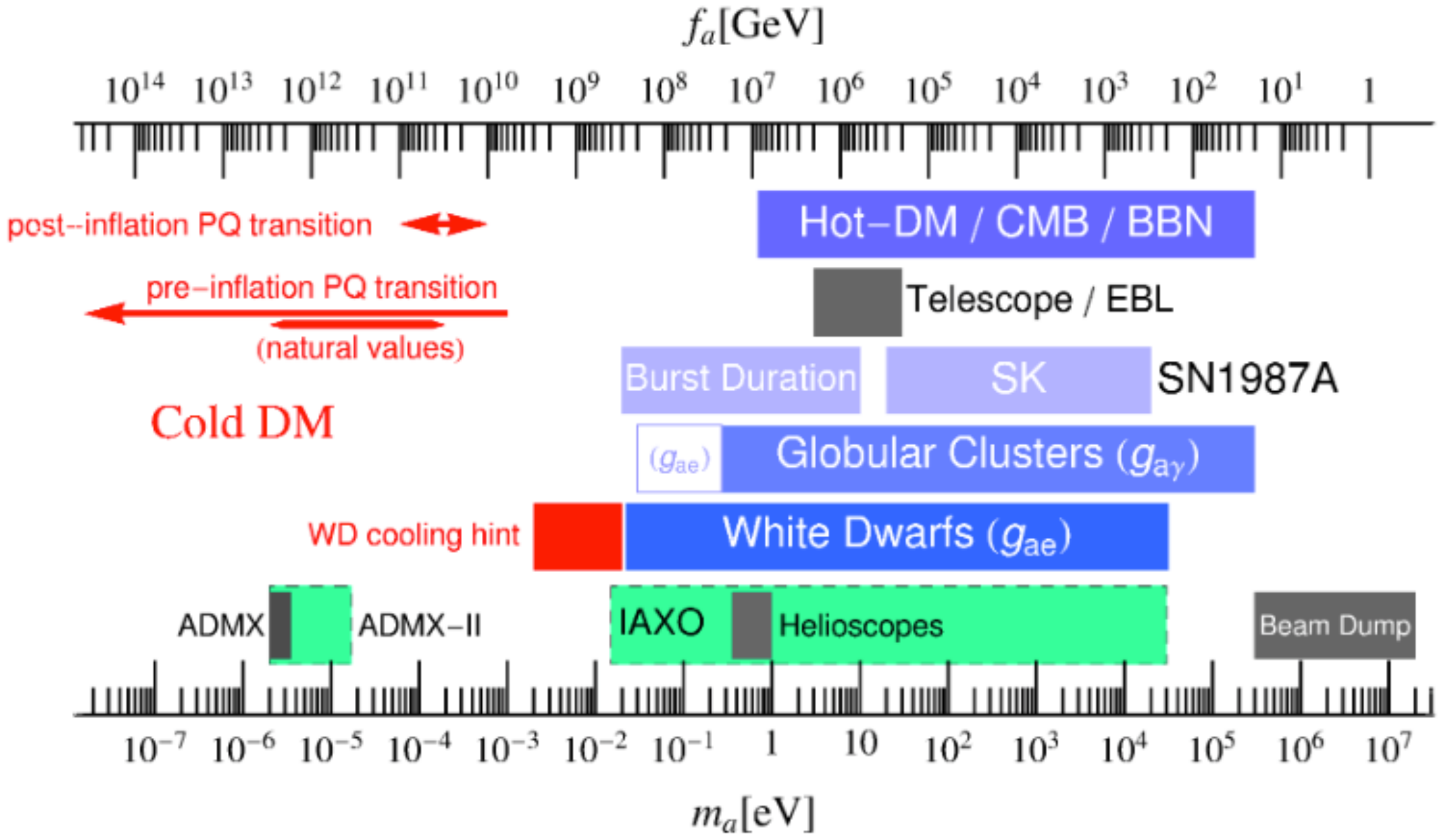}
\caption{ \label{Fig:1} Limits on the axion mass established by cosmology and astrophysics.  Light grey regions are very model dependent. (From Ref. \cite{Agashe:2014kda}.)}
\end{center}
\end{figure}

The range $10^{- (6-2)}$ eV is traditionally been regarded as the open mass window for axions; see Figure \ref{Fig:1} \cite{Agashe:2014kda}.  However, as previously mentioned, string theories are replete with axions or axion-like particles, upwards of a hundred within any particular realization, but such theories naturally favor $f_a \approx 10^{( 15-16)} \, \text{GeV}$, corresponding to neV-scale masses.  It is impossible to say which, if any of these  solve the Strong-CP problem, and which, if any would be cosmologically significant.  String theory axions and possible limits from isocurvature fluctuations will be reviewed briefly in Sec. 3.4 as a preamble to the discussion of the NMR-based experiment, CASPEr.

\subsection{Cavity microwave experiment}
Problematically such light axions would be so weakly coupled as to be undetectable in conventional experiments.  In 1983, Pierre Sikivie resolved this conundrum by showing that axions constituting the Milky Way halo could resonantly convert into a monochromatic microwave signal in a high-Q microwave cavity permeated by a strong magnetic field \cite{Sikivie:1983ip, Sikivie:1985yu}, with the conversion power given by
\begin{equation}
P_\text{SIG} = \eta g_{a \gamma \gamma}^2 \left( \frac{\rho_a}{m_a} \right) B_0^2 V C Q_L
\end{equation}

The physics parameters, beyond the control of the experimentalist, are the axion-photon coupling constant $g_{a \gamma \gamma}$, the axion mass $m_a$, and the local density of axions in the halo, $\rho_a$.  Within experimental control are the magnetic field strength $B_0$, and the volume of the cavity $V$, as well as the mode-dependent form-factor $C$, and loaded quality factor of the cavity $Q_L$, i.e. the quality factor with power coupled out to the receiver.  $\eta$ is the fraction of power coupled out by the antenna probe, generally adjusted to be at or near critical coupling, $\eta = 1/2$.  The resonant conversion condition is that the frequency of the cavity must equal the mass of the axion, $h \nu = m_a c^2 \left[ 1 + \frac{1}{2} \mathcal{O}(\beta^2) \right]$, where $\beta \approx 10^{-3}$ is the galactic virial velocity.  The signal is thus monochromatic to $10^{-6}$.  The search is performed by tuning the cavity in small overlapping steps (Figure \ref{Fig:2}).

\begin{figure}
\begin{center}
\includegraphics[width= \columnwidth]{Fig2.pdf}
\caption{ \label{Fig:2} Schematic of the microwave cavity search for dark matter axions. Axions resonantly convert to a quasi-monochromatic microwave signal in a high-Q cavity in a strong magnetic field; the signal is extracted from the cavity by an antenna, amplified, mixed down to the audio range, and the power spectrum calculated by a FFT. 
%The TM$_\text{010}$ is by far the optimal mode for the experiment; the cavity can be tuned $\pm 50\%$ in frequency with metal or dielectric rods. 
Possible fine structure on top of the thermalized axion spectrum would reveal important information about the formation of our galaxy.}
\end{center}
\end{figure}

The expected signal power is extraordinarily tiny, of order $10^{-22}$ W for the current experiment.  Actual detection of the axion is the consummate signal-processing problem, governed by the Dicke radiometer equation \cite{Dicke}
\begin{equation}
\label{eqn: dicke}
\frac{S}{N} = \frac{P_\text{SIG}}{k T_\text{SYS}} \sqrt{\frac{t}{\Delta \nu}},
\end{equation}
where $S/N$ is the signal to noise ratio, and the total system noise temperature $T_\text{SYS} = T+ T_N$ is the sum of the physical temperature $T$ and the intrinsic amplifier noise temperature $T_N$, with $k$ the Boltzmann constant.  The integration time is $t$, and the bandwidth of the axion signal is $\Delta \nu$, where it is assumed that the resolution of the spectral receiver is much better than the width of the axion signal.  

One especially important feature about the microwave cavity search for axions that strongly differentiates it from WIMP searches, is that it is a total energy detector, i.e.~the signal represents the instantaneous (mass + kinetic) energy of the axion.  While the majority of the signal strength will almost certainly be found in a broad quasi-Maxwellian distribution of width $\frac{\Delta \nu}{\nu} \sim 10^{-6}$, there has been a great deal of speculation and research over the past two decades about the phase space structure of the axion signal, caustics, fine structure due to late infall axions, etc. \cite{Sikivie:2010bq}.  A high resolution channel has been implemented on ADMX based on a Fast Fourier Transform (FFT) of an entire subspectrum, the basic unit of data collection, which can resolve structure down to the transform limit, e.g. $\frac{\Delta \nu}{\nu} \sim 10^{-11}$ for a 100-second run at 1 GHz.  Due to the motion of the laboratory through the dark-matter halo ($v_\text{ROT} \sim 0.4 \, \text{km}/\text{sec}$, $v_\text{ORB} \sim 30 \, \text{km}/\text{sec}$), any fine structure would exhibit both diurnal and sidereal modulations in frequency.
%; in fact a peak of transform-limited width would shift in frequency by its FWHM in a single 100-second run.  Figure 3 shows the frequency modulation of four flows from one representative infall model \cite{Sikivie:1998dr}, at the latitude of the experiment \cite{vanBibber:2003sv}.
%As the earth's spin is not co-linear with its orbital rotation, the net motion of the laboratory through the halo spans the full 3-dimensional basis in velocity over time.
A little reflection makes it clear that should such fine structure be found, fitting the amplitude and phase of the diurnal and sidereal oscillation in frequency would uniquely determine each vector flow to high precision, truly opening the field of dark matter astronomy \cite{vanBibber:2003sv}.  While N-body simulations strongly support a hierarchical and chaotic picture of structure formation, these simulations themselves exhibit significant mesoscale substructure in phase space, which could be studied.	

%\begin{figure}
%\begin{center}
%\includegraphics[width= \columnwidth]{Fig3.pdf}
%\caption{ \label{Fig:3} Modulation in frequency of four flows from one particular Milky Way infall model of Sikivie \cite{Sikivie:1999tn}.}
%\end{center}
%\end{figure}

\subsection{ADMX and ADMX-HF}

\subsubsection{Early experiments}

\begin{figure}
\begin{center}
\includegraphics[width= \columnwidth]{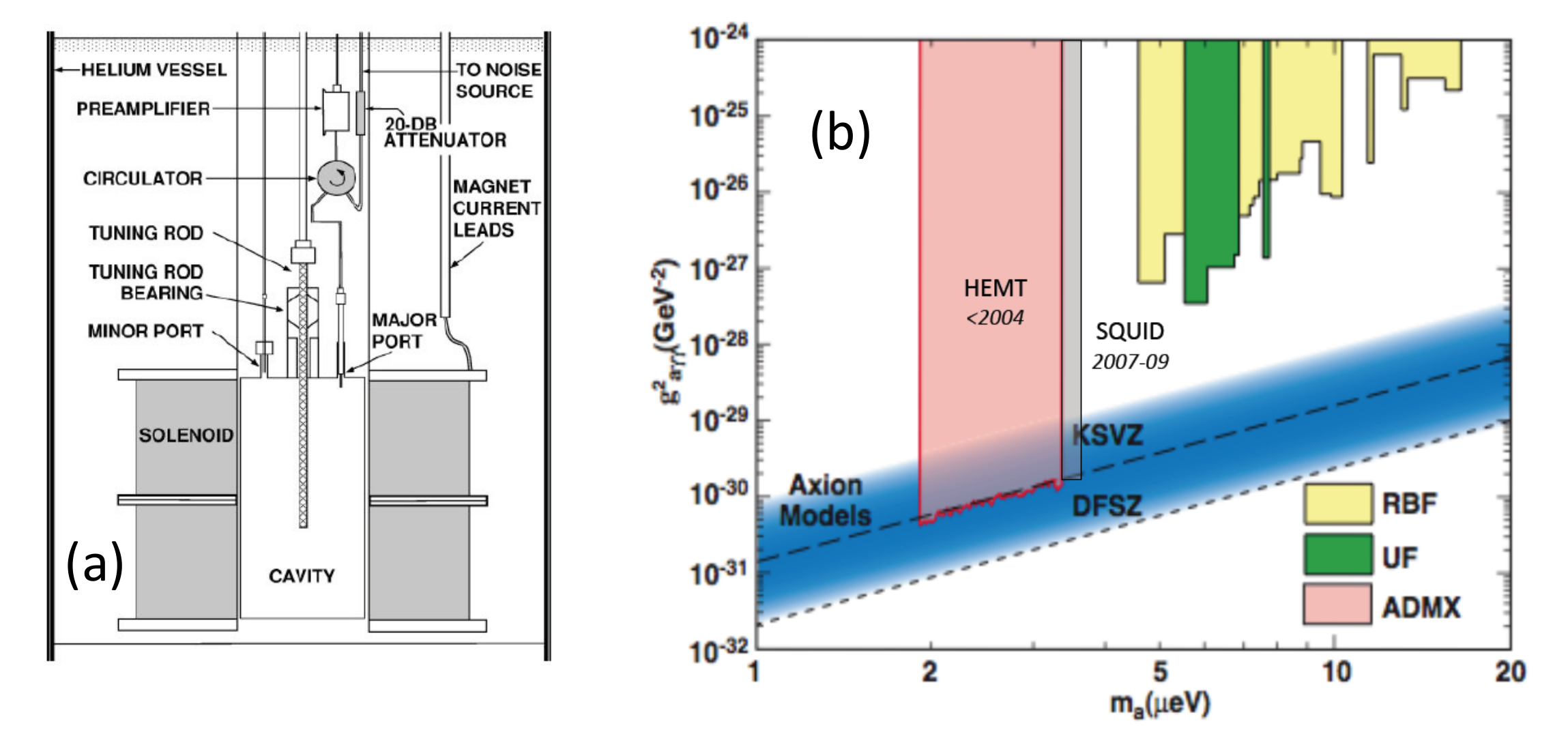}
\caption{ \label{Fig:4} (a) Layout of the Rochester-Brookhaven-Fermilab (RBF) experiment \cite{DePanfilis:1987dk, Wuensch:1989sa}.  (b) The 90\% c.l. exclusion regions for the RBF, UF and ADMX experiments.}
\end{center}
\end{figure}

Two pilot efforts in the 1-4 GHz range, at BNL \cite{DePanfilis:1987dk, Wuensch:1989sa} and the University of Florida \cite{Hagmann:1990tj}, were mounted soon after publication of the experimental concept.  These employed cavities of a few liters volume, and the best conventional amplifiers at that time, e.g. Heterojunction Field Effect Transistor (HFET) amplifiers. With noise temperatures only in the $T_N \sim$ 3-20 K range, they did not have the sensitivity to reach Peccei-Quinn axions (Figure \ref{Fig:4}).  Nevertheless, these two experiments developed much of the design philosophy and know-how about microwave cavities that the current experiments still build on today.

The CARRACK experiment in Kyoto marked another significant development in the history of the microwave cavity experiment.  The goal of CARRACK was to achieve a dramatic reduction in the system noise temperature, both by reducing the physical temperature of the experiment down to $\sim 15$ mK with a $^3$He-$^4$He dilution refrigerator, and by utilizing a Rydberg-atom single-quantum detector in lieu of a standard linear amplifier to eliminate the amplifier noise contribution.  Linear amplifiers are ultimately subject to an irreducible noise contribution, the Standard Quantum Limit, $kT_{SQL} = h \nu$.   The Rydberg-atom single-quantum detector can effectively be thought of as a tunable ``RF photomultiplier tube", for which the photon interacts as a particle rather than a wave, thus circumventing the SQL. Tada et al. were able to measure the the blackbody photon spectrum of the cavity at 2.527 GHz as a function of temperature all the way down to $T = 67$ mK, nearly a factor of two below the Standard Quantum Limit of $T_{SQL} \sim$ 120 mK \cite{Tada2006488}.  From the technical perspective, CARRACK was successful, but ultimately proved too complex to be feasible as a production experiment.

\subsubsection{Axion Dark Matter eXperiment (ADMX)}

Drawing on the experience from the Rochester-Brookhaven-Fermilab (RBF) and the University of Florida (UF) searches, the ADMX collaboration designed an experiment with the goal to reach KSVZ axions saturating our galactic halo, whose local density would be $\rho_a \sim 0.45 \, \text{GeV}/\text{cm}^3$.  This goal would be achieved both by a scale-up of the cavity volume by two orders of magnitude, and profiting from the steady improvement in the noise temperature of commercial HFET amplifiers.  

\begin{figure}
\begin{center}
\includegraphics[width= \columnwidth]{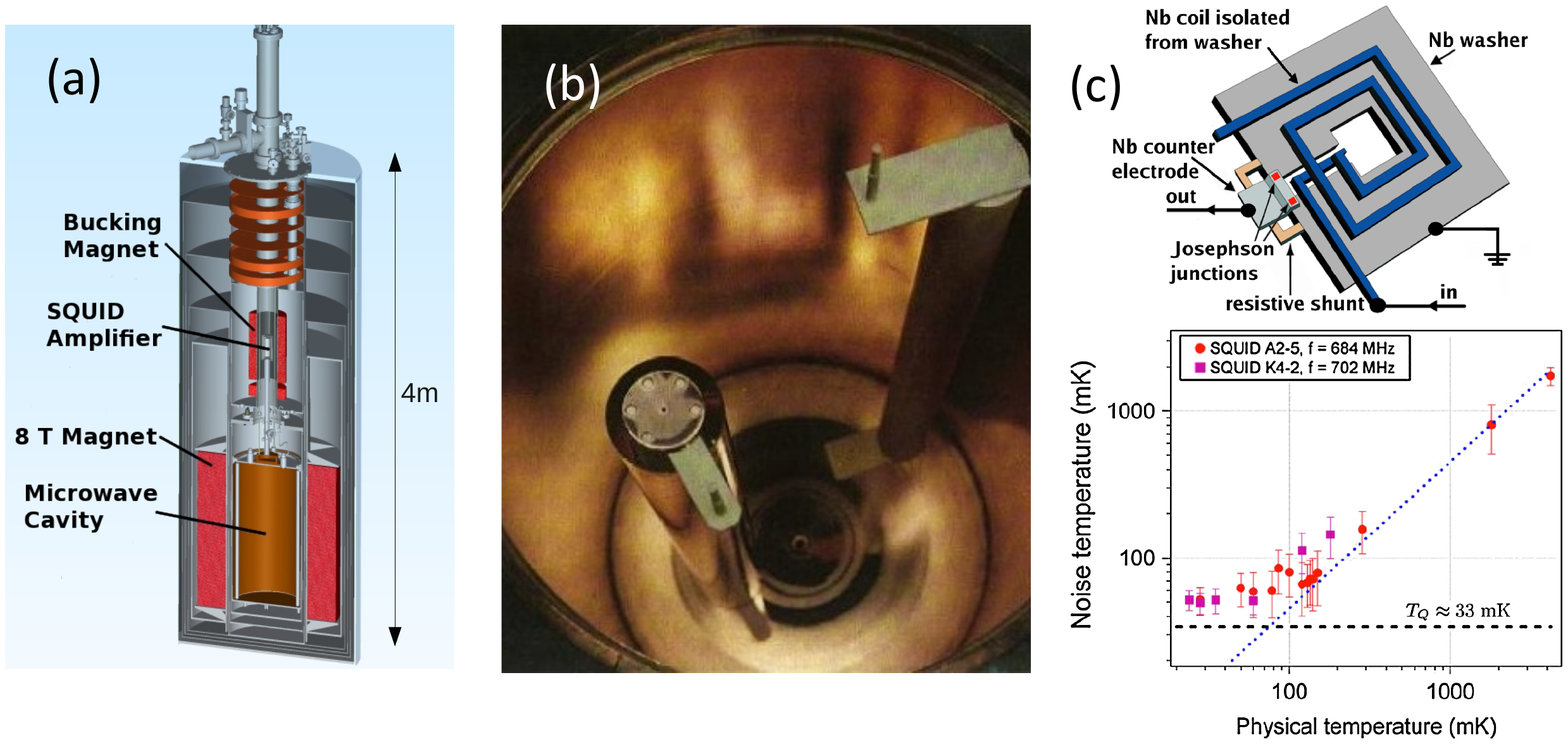}
\caption{ \label{Fig:5} The Axion Dark Matter eXperiment.   (a) Schematic layout.  (b) Microwave cavity and tuning rods.  (c) dc SQUID amplifiers.  In addition to being near-quantum limited, the MSAs have been demonstrated to be tunable, work with a reactive load, and can be staged \cite{Muck, Muck:1999nra, Muck2}.}
\end{center}
\end{figure}

The NbTi superconducting magnet has an inner bore 60 cm $\times$ 110 cm, and sustains a maximum central field of 8 T.
%It is built into a 3.4 m high cryostat, the total magnet assembly weighing 11 tons.
The microwave cavities are made by electrodepositing high-purity copper on a stainless steel body, followed by annealing (Figure \ref{Fig:5}), leading to cavity quality factors of $Q \sim 10^5$;
%By axially translating a large metal or dielectric rod between the wall and the center of the cavity, the TM$_{010}$ mode of the cavity can be shifted upward or downward by roughly 30\%; 
the experiment is  tuned in small overlapping steps of the cavity bandpass. To date, the experiment has been cooled to superfluid helium temperatures, $T \sim 1.5$ K.  

For the first operational phase of ADMX (1995-2004), HFET amplifiers made by NRAO were used, ultimately reaching a noise temperature $T_N \sim $1.5 K, with the system noise temperature thus being $T_{SYS} \sim$ 3 K.  For the second operational phase (2007-09), Microstrip-coupled SQUID Amplifiers (MSA) developed specifically for ADMX were employed, whose noise temperature was demonstrated on the bench to be $T_N < 1.5 \, T_{SQL}$ at frequencies up to a GHz, when cooled to 30 mK \cite{Muck, Muck:1999nra, Muck2}.  However, the noise of these MSAs exhibits a strong temperature dependence, and at pumped helium temperatures, are not better than transistor-based amplifiers; see Figure \ref{Fig:5}.  Nevertheless, demonstrating that dc SQUID amplifiers could be made to work successfully in situ represented a great advance for the experiment, and a dilution refrigerator will be incorporated into the experiment in 2015, enabling the experiment to achieve $T_{SYS} < 200$ mK, sensitive to DFSZ axions even with less than saturation density.  

To date, ADMX has covered 460 - 890 MHz in frequency (1.9 - 3.65 $\mu$eV) or slightly less than an octave in mass range \cite{Asztalos:2009yp, Hoskins:2011iv}; Figure \ref{Fig:4}.  This underscores the importance both of concurrent R$\&$D on new cavity and amplifier concepts for frequencies much greater than a GHz, as well as greatly increasing the scanning rate of the experiment.  

\subsubsection{ADMX-HF (High Frequency)}

\begin{figure}
\begin{center}
\includegraphics[width= \columnwidth]{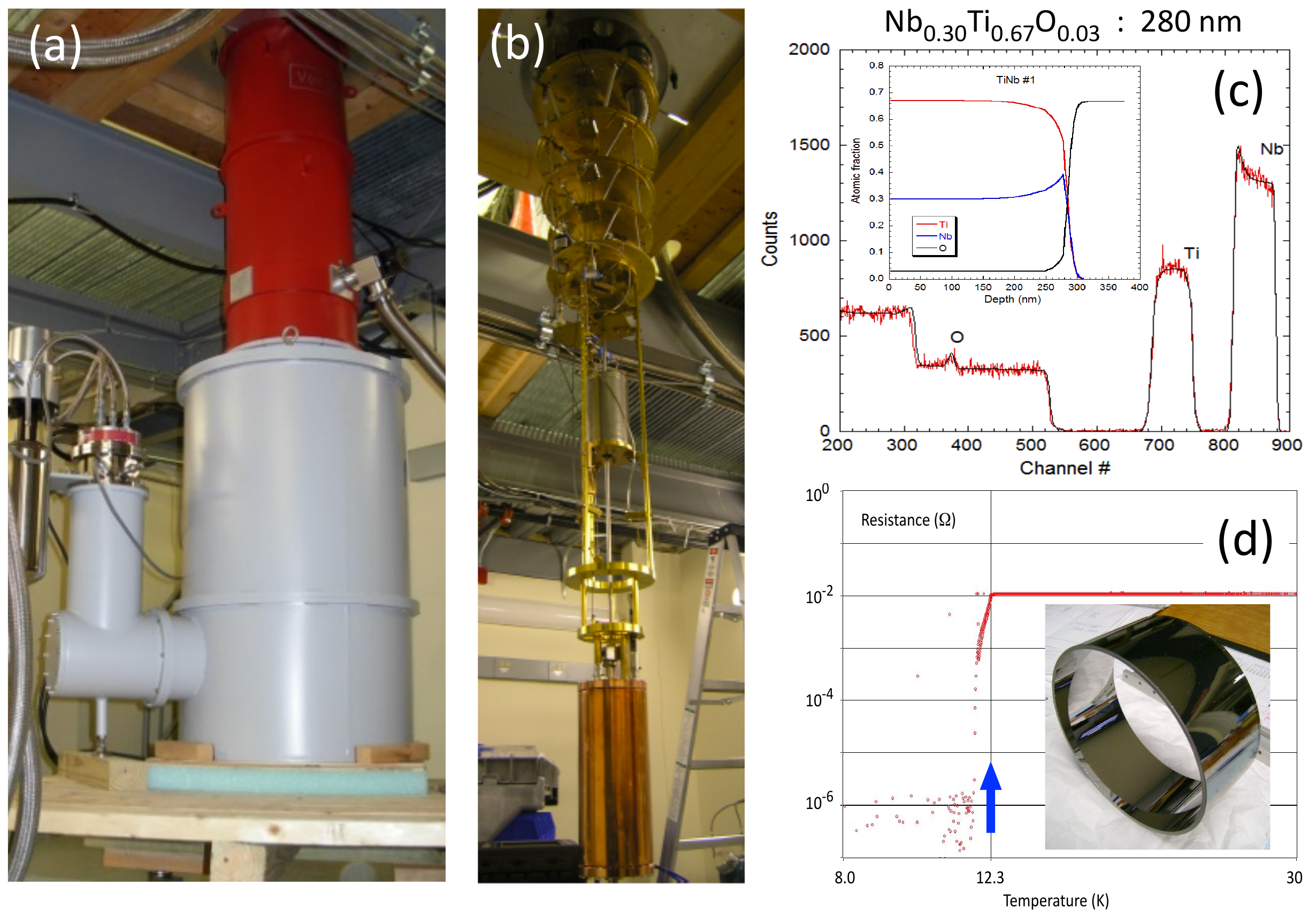}
\caption{ \label{Fig:6} (a) ADMX-HF below deck, showing the magnet (grey) and dilution refrigerator (red).  (b) Experimental gantry, showing the dilution refrigerator (top), the JPA magnetically shielded canister (middle) and microwave cavity (bottom).  (c) Rutherford backscattering profile of a thin-film superconductor.  (d) Resistance vs. temperature, demonstrating $T_C > 12$ K.  (Insert)  Thin-film superconductor deposited on the inside of a 10 cm diameter quartz tube.}
\end{center}
\end{figure}

A second smaller ADMX platform has been constructed and commissioned at Yale, precisely to develop technologies and techniques applicable to the next higher decade in mass \cite{Shokair:2014rna}.  The NbTi solenoidal magnet (9.4 T,  60 cm $\times$ 110 cm) was designed for exceptionally high field uniformity ($B_r < 50$ G) for the purposes of testing thin-film superconducting cavities, although initially the cavity is electrodeposited copper (Figure \ref{Fig:6}). ADMX-HF incorporated a dilution refrigerator from the outset, resulting in a base temperature of the experiment $T \sim 25$ mK. Josephson Parametric Amplfiers (JPA) are well suited to the initial 5 GHz range of the experiment, where they possess high gain (20-30 dB), are tunable over an octave and can operate with a system noise temperature $T_N \sim T_{SQL}$ \cite{CastellanosLehnert, Castellanos}; this noise was achieved within a factor of two in its first commissioning run.  In its initial configuration, the experiment is projected to reach a sensitivity in axion-photon coupling $\sim 2 \, \times$ KSVZ.

Two promising lines of R$\&$D are currently being pursued with high priority.  The first is the prospect of incorporating Type-II superconducting thin films on all cylindrical surfaces of the microwave cavity to boost the Q by an order of magnitude, to which the axion-photon conversion power and thereby scanning rate are directly proportional.  Xi et al. have recently demonstrated that very thin films of Nb$_x$Ti$_{1-x}$N are perfectly microwave reflecting to frequencies $>$ 100 GHz, immersed a magnetic field parallel to its surface up to $B_{||} =$ 10 T \cite{Xi}. Encouraged by this report, a program has been initiated of making and characterizing thin (10-200 nm) NbTiN films by RF plasma deposition.  Satisfactory films on planar samples were readily achieved that are non-critical in exact composition (Figure \ref{Fig:6}).  RF tests of small prototype cavities will be conducted next, and finally testing them in a magnetic field.  Successful implementation in ADMX-HF will require minimizing lossy flux vortices penetrating the superconducting surfaces; this requires the film be both extremely thin, and  parallel to the magnetic field.  To this end, the magnet was designed to meet strict specifications on field uniformity to minimize radial components.
% of the field, and the cavity will likewise need a precision alignment system on the gantry.  

The other major R$\&$D effort will be the incorporation of a receiver based on squeezed-vacuum states to evade the quantum limit in noise, employed so far only by the laser interferometric gravity wave community, i.e. LIGO and GEO.   The JILA group in ADMX-HF have used one JPA to measure the squeezed noise generated by a second JPA \cite{Mallet} %\cite{Lamoreaux:2013koa}, 
achieving an equivalent noise temperature $T_N = h \nu /4$.  Indeed, this experiment can be regarded as a proof-of-principle demonstration of a quantum-noise evading axion search. Not only was it possible to measure noise in one quadrature below the standard quantum limit, but also to demonstrate that quadrature had fluctuations below vacuum because it had been squeezed.
%Taking this receiver from the bench into the environment of the microwave cavity experiment will require significant reengineering, including the conversion from coax cable to rigid waveguide to eliminate transmission losses.  The effort would be worthwhile as it is projected that with care, 
The system noise could conceivably be reduced to 1/10 of the quantum limit.  Together with the gain associated with the superconducting thin-film cavity, this small-volume experiment could ultimately reach DFSZ sensitivity.   

Examining quantum noise in the cavity and amplifier more rigorously, Lamoreaux et al. have concluded that at higher frequencies, and thus higher axion masses, single-photon detectors become competitive and ultimately favored, when compared to quantum-limited linear amplifiers, as the detector technology in the microwave cavity experiment \cite{Lamoreaux:2013koa}, and that the cross-over point in this comparison is not far above the frequencies of the current ADMX-HF search range, of order 10 GHz.  The microwave cavity experiment can profitably leverage advances in superconducting qubit readout schemes \cite{Schuster, Wallraff, Chen}.

\subsubsection{New Directions and Concepts}

The past several years have seen a number of new search concepts for the detection of dark matter axions.  The NMR-based experiment CASPEr to explore the 10$^{-(6-9)}$ eV mass range will be discussed separately in Sec. 3.4.  However  other promising innovations warrant brief mention here.

Rybka et al. have described an experiment designed on an open microwave resonator structure appropriate for axions in the 100 -1000 $\mu$eV decade, where the wavelength of the microwave photon after conversion ranges from $\lambda =$ 10 - 1 mm \cite{Rybka:2014cya}.  As the conversion probability maximizes strongly when the applied magnetic field B(r) follows the electric field of the photon mode E(r), a wiggler-like magnetic field is required with a continuously tunable periodicity to sweep out the range of masses.  Rybka  et al. effect this with a series of wire planes in the open resonator of alternating sign in current; ultimately these planes will consist of superconducting wires or stripes patterned on a thin substrate.  A first prototype was able to exclude axions as the dark matter between 68.2 and 76.5 $\mu$eV, with couplings $g_{a\gamma\gamma} > 4 \times 10^{-7} \text{GeV}^{-1}$, not far removed from existing limits from laser-based limits, remarkably with a magnetic field of $|B| < 10$ gauss.  Ultimately sensitivity to couplings of $10^{-15}$ GeV$^{-1}$ should be achievable, below DFSZ axions for that mass range.
%To date, this proposal perhaps represents the only feasible path by which to explore this critical mass range.

%\begin{figure}
%\begin{center}
%\includegraphics[width= \columnwidth]{Fig7.pdf}
%\caption{ \label{Fig:7}  (a) Concept of the Orpheus experiment to search for dark-matter axions in an confocal resonator configuration \cite{Rybka:2014cya}, appropriate for masses in the 10$^{-(3-4)}$ eV.  (b)  An experiment based on a discrete LC circuit in a magnetic field, filling an important gap in the 10$^{-(7-9)}$ eV range \cite{Sikivie:2013laa}.}
%\end{center}
%\end{figure}

Sikivie  et al. have focused on the other challenge of the microwave cavity experiment, namely extending the search downward in mass  \cite{Sikivie:2013laa}, where much below $10^{-6}$ eV ($\sim 250$ MHz) the cavity and the magnet which encloses it become unfeasibly large.
%The frequency of the TM$_{010}$ mode is inversely proportional to the cavity diameter, thus the lower limit in mass is now determined by the largest feasible diameter magnet inside of which the cavity must fit; to extend the search significantly below 10$^{-6}$ eV with a conventional cylindrical would require a very large and expensive magnet.
What they propose to circumvent this limitation is replacing the cavity with a lumped-parameter LC circuit external to the magnetic field, excited by a pickup loop inside the magnet  threaded by the transverse B-field of the mode.  Optimistically, the authors represent that with a magnet such as that of the the current ADMX, this technique would be maximally sensitive around 10$^{-7}$ eV, probing the band of Peccei-Quinn axion models.  If the concept can be fully developed and implemented, such a lumped-parameter LC circuit search for dark matter axions could bridge a key gap between where the cavity-based experiment and the NMR-based experiment may have difficulty in overlapping.

Other new and imaginative concepts have been put forward recently.  These include instrumenting magnets of extreme aspect ratio, e.g. high energy dipoles, with microwave cavities \cite{Baker:2011na} as well as a proposal for a dish-geometry resonator that would be both directional and broadband in frequency \cite{Horns:2012jf}.  Such a dish antenna in fact has already been implemented for a Hidden Photon dark matter search.

\subsection{Detection with NMR: CASPEr}
\label{section: CASPEr}

The Cosmic Axion Spin Precession Experiment (CASPEr) proposal aims to open a new direction in the experimental search for axion dark matter.  CASPEr will detect the spin precession caused by axion dark matter using nuclear magnetic resonance (NMR) techniques \cite{Budker:2013hfa, Graham:2013gfa}. This novel approach complements existing efforts: ADMX is sensitive to the higher axion frequencies, whereas CASPEr will cover the lower frequencies where the axion arises from energies $f_a \sim 10^{15} \, \text{GeV} - 10^{19} \, \text{GeV}$\footnote{It used to be argued that this range was disfavored by cosmology, but that requires specific assumptions about initial conditions which are easily violated, thus this range is allowed, see e.g.~\cite{Linde:1987bx, Wilczek:2012it}, and is also well-motivated theoretically \cite{Svrcek:2006yi}.}.  This range is very challenging for any other technique to reach, though some astrophysical techniques may be able to probe it \cite{Arvanitaki:2014wva, Arvanitaki:2010sy, Arvanitaki:2009fg}.   A detection in such an experiment would not only represent the discovery of dark matter but would also provide insights into the high-energy scales from which such an axion would arise, near fundamental scales such as the grand unification, Planck, or string scales.

Almost all axion experiments search for the coupling of the axion to photons.  CASPEr searches for two different couplings of the axion and thus naturally divides into two experiments: CASPEr-Wind and CASPEr-Electric.  The Wind experiment searches for the `axion wind' effect, the direct coupling of the axion to the spin of the nucleus \cite{Graham:2013gfa}.  This is the pseudoscalar coupling 
\begin{equation}
\mathcal{L} = ... + g_\text{aNN} \left( \partial_\mu a \right) \bar{N} \gamma^\mu \gamma_5 N
\label{eqn:gaNN}
\end{equation}
which physically causes a precession of a nucleon spin around the spatial gradient of the local axion dark matter field \cite{Graham:2013gfa}.
CASPEr-Electric searches for the time-varying nucleon electric dipole moment (EDM) caused by the axion \cite{Budker:2013hfa}, which can be written as the coupling of the axion to nucleons
\begin{equation}
\label{eqn: axion EDM coupling}
\mathcal{L} = ...  -\frac{i}{2} g_d \, a \, \bar{N} \sigma_{\mu \nu} \gamma_5 N F^{\mu \nu}.
\end{equation}
where $F$ is the field strength of electromagnetism.
This coupling arises from the fundamental defining coupling of the QCD axion to gluons $\propto \frac{a}{f_a} G \tilde{G}$ \cite{Graham:2011qk}.  Both of these effects are time-varying because the background axion dark matter field $a$ oscillates at a frequency equal to its mass.  The CASPEr idea could also be used to search for the coupling of the axion to electron spin but that does not appear sensitive enough to get beyond current limits in that parameter space \cite{Graham:2013gfa}.  There has been a long history and significant recent interest in looking at such effects on nucleons and electrons \cite{Stadnik:2013raa, Stadnik:2014ala, Hong:1991fp}.

\begin{figure}
\begin{center}
\includegraphics[width=3.5 in]{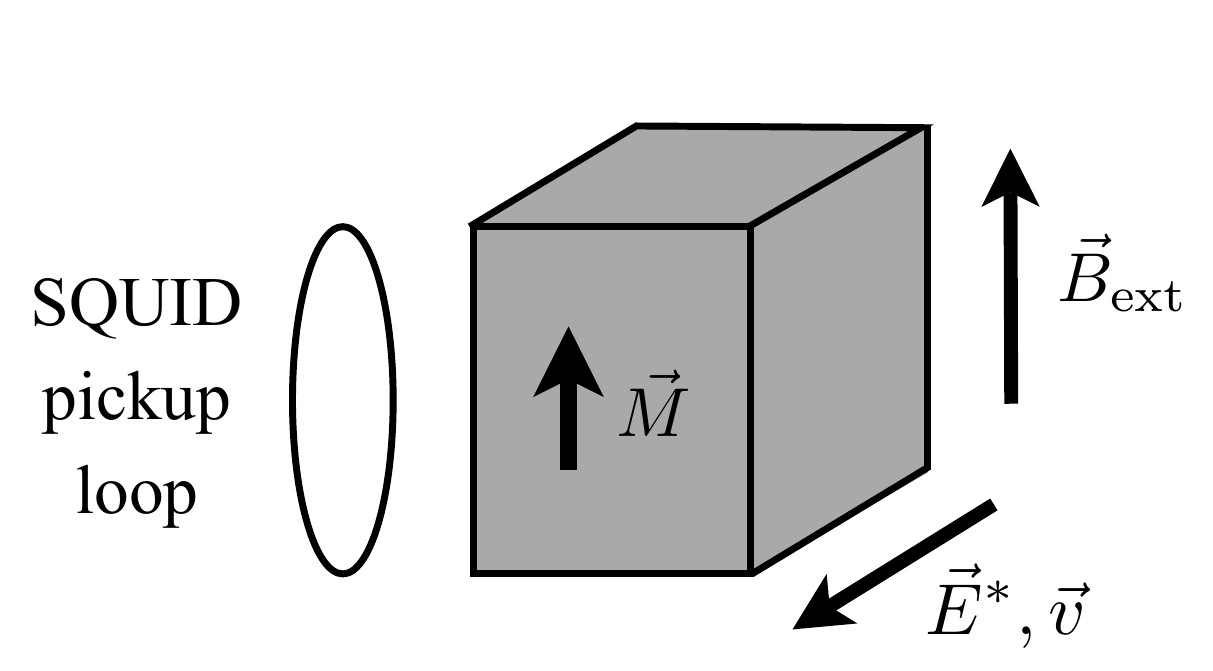}
\caption{ \label{Fig:CASPErsetup} CASPEr setup.  The applied magnetic field $\vec{B}_\text{ext}$ is colinear with the sample magnetization, $\vec{M}$.   In CASPEr-Wind the nuclear spins precess around the local velocity of the dark matter, $\vec{v}$, while in CASPEr-Electric the nuclear EDM causes the spins to precess around an effective electric field in the crystal $\vec{E}^*$, perpendicular to $\vec{B}_\text{ext}$.  The SQUID pickup loop is arranged to measure the transverse magnetization of the sample.}
\end{center}
\end{figure}

The main idea behind CASPEr is to use the time-varying nature of the effect (either EDM or wind) to cause precession of nuclear spins in a sample of material.
The Larmor frequency of the nuclear spins is scanned by ramping the magnetic field and at the frequency corresponding to the mass of the axion an NMR signal is observed in the usual way using a precise magnetometer as in Figure \ref{Fig:CASPErsetup}.
%An external applied magnetic field is scanned over and when the Larmor frequency of the nuclear spins matches the axion frequency, a resonant enhancement occurs analogous to NMR.  Precession of the sample spins then leads to a transverse magnetization which can be measured with a precise magnetometer as in Figure \ref{Fig:CASPErsetup}.  
There are many choices of possible sample material which are beginning to be tested experimentally and explored theoretically \cite{Roberts:2014cga, Roberts:2014dda} to find the optimum.

\begin{figure}
\begin{center}
\includegraphics[width=\columnwidth]{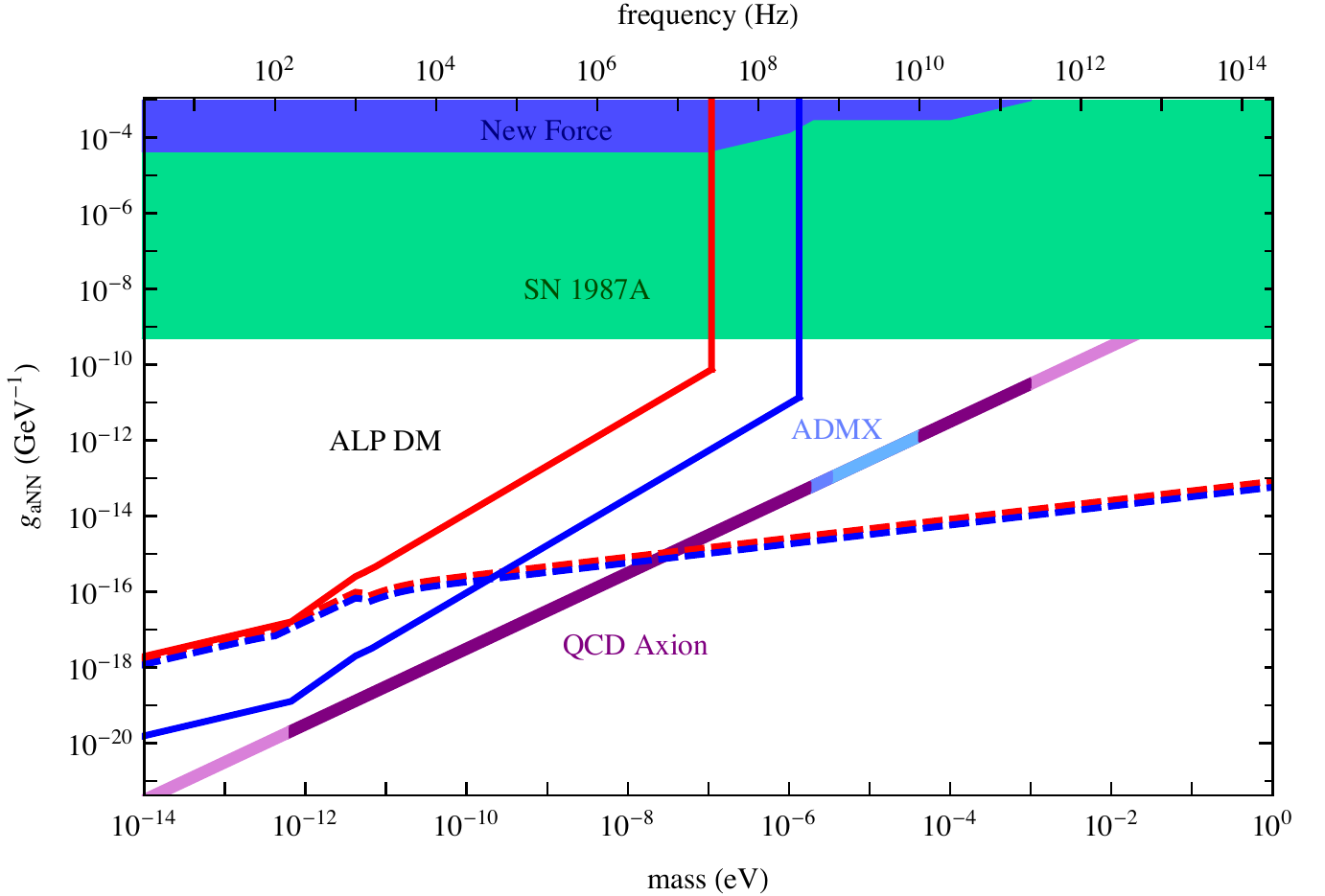}
\caption{ \label{Fig:CASPErWIND} Sensitivity of the CASPEr-Wind proposal.  ALP parameter space in pseudoscalar coupling of axion to nucleons Eqn.~\ref{eqn:gaNN} vs mass of ALP.  The purple line is the region in which the QCD axion may lie.  The width of the purple band gives an approximation to the axion model-dependence in this coupling.  The darker purple portion of the line shows the region in which the QCD axion could be all of the dark matter and have $f_a < M_\text{pl}$ as in Figure \ref{Fig:EDM}.  The green region is excluded by SN1987A from \cite{Raffelt:2006cw}. The blue region is excluded by searches for new spin dependent forces between nuclei. The red line is the projected sensitivity of an NMR style experiment using Xe, the blue line is the sensitivity using $^3\text{He}$.  The dashed lines show the limit from magnetization noise for each sample. The ADMX region shows the part of QCD axion parameter space which has been covered (darker blue) \cite{Asztalos:2009yp}  or will be covered in the near future (lighter blue) by ADMX.  For full details see  \cite{Graham:2013gfa}.}
\end{center}
\end{figure}

\begin{figure}
\begin{center}
\includegraphics[width=\columnwidth]{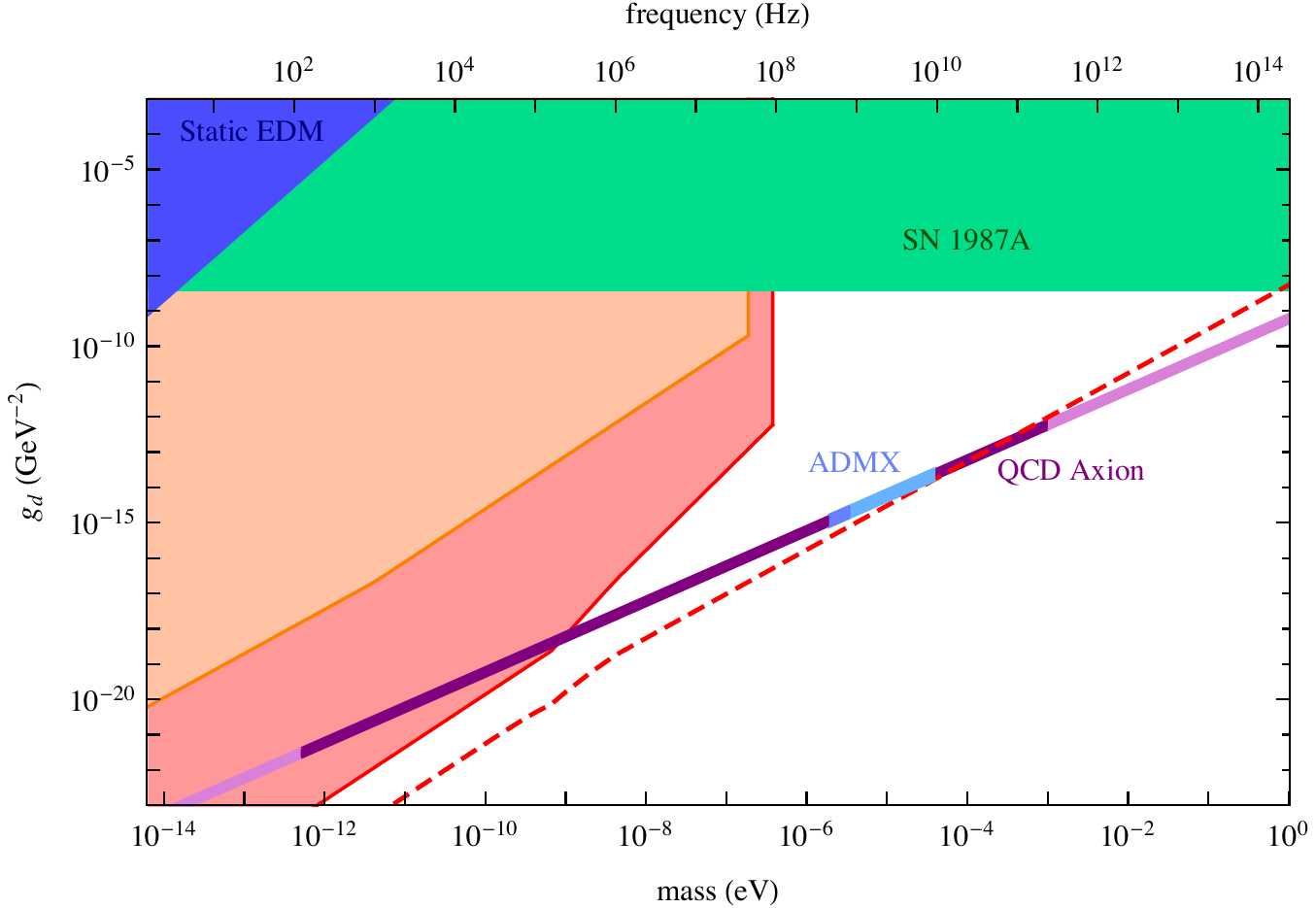}
\caption{ \label{Fig:EDM}  Sensitivity of the CASPEr-Electric proposal.  Estimated constraints in the ALP parameter space in the EDM coupling $g_d$ (as in Equation \ref{eqn: axion EDM coupling}) vs.~the ALP mass.
%The green region is excluded by the constraints on excess cooling of supernova 1987A \cite{ALP space}.
The blue region is excluded by existing, static nuclear EDM searches \cite{Graham:2013gfa}.
%The QCD axion is in the purple region, whose width shows the theoretical uncertainty \cite{ALP space}.
The solid red and orange regions show projected sensitivity estimates for CASPEr-Electric phase 1 and 2 proposals, set by magnetometer noise.  The red dashed line shows the limit from magnetization noise of the sample for phase 2.
Other regions are as in Figure \ref{Fig:CASPErWIND}.  For full details see \cite{Budker:2013hfa, Graham:2013gfa}.
%The ADMX region shows what region of the QCD axion has been covered (darker blue) \cite{Asztalos:2009yp} or will be covered (lighter blue) \cite{ADMXwebpage, snowdarktalk}.
%Phase 1 is a modification of current solid state  static EDM techniques that is optimized to search for a time varying signal and can immediately begin probing the allowed region of ALP dark matter.  To calculate limits from previous (static) EDM searches as well as our sensitivity curves, we assume the ALP is all of the dark matter.
%Current EDM techniques that are optimized to search for a time-varying EDM can already search for ALP dark matter in the allowed region of parameter space \cite{ALP space}.
}
\end{center}
\end{figure}

The Wind experiment is technically simpler since it does not require an applied electric field and hence can be done using LXe as the sample for which the required NMR techniques have already been perfected.  It can cover large parts of general axion (or axion-like particle) parameter space, many orders of magnitude beyond current constraints as in Figure \ref{Fig:CASPErWIND}. It also provides a stepping stone towards the CASPEr-Electric experiment, as within the Wind experiment, many of the key technologies needed for the Electric experiment will be developed. CASPEr-Electric requires a more complicated material such as a ferroelectric or polar crystal with a large internal electric field, but it has a better ultimate sensitivity, allowing it to reach all the way to the QCD axion over several orders of magnitude in frequency space that are unreachable by other techniques as in Figure \ref{Fig:EDM}\footnote{Note that the Wind coupling leads to a spin-dependent force which could be probed using NMR techniques as well e.g.~\cite{Moody:1984ba, Vasilakis:2008yn, Burghoff:2011zz, Ledbetter:2012xd, Heil:2013tpa, Tullney:2013wqa, Arvanitaki:2014dfa}.}.

The CASPEr-Wind experiment is in fact a search for any light particle that couples to nuclear spin (a generic coupling), not just the axion.  For example, any pseudo-Goldstone boson is expected to possess a coupling that would be detectable in the CASPEr-Wind experiment.  It can also detect other types of dark matter, for example hidden photon dark matter \cite{Nelson:2011sf, Arias:2012az} is detectable through a nuclear dipole moment coupling.

Existing experiments may already be able to set limits on axion-like particles.  Data from experiments searching for nuclear EDMs or looking at nucleon spin precession in a low background environment may be reanalyzed to search for a time-varying signal, a sign of the axion.  While not ultimately as sensitive as CASPEr where the signal is resonantly enhanced, such searches may be able to probe beyond the current astrophysical limits in Figures \ref{Fig:CASPErWIND} and \ref{Fig:EDM}.

CASPEr is a novel and highly sensitive search for a broad class of dark matter candidates in two new parameter spaces, the `axion wind' and nuclear EDM, of which the QCD axion is the most well-known example.
%CASPEr is sensitive to a broad class of new light particles in two new parameter spaces, the `axion wind' and nuclear EDM.
In particular, CASPEr has the sensitivity to detect the QCD axion over a wide range of masses from $\sim 10^{-9}$ eV to $10^{-12}$ eV which are well-motivated by fundamental physics \cite{Svrcek:2006yi} and where no other experiment can detect it.

Construction is just beginning on the CASPEr experiment.  Work on CASPEr is currently being carried out in several places including Stanford, Berkeley, and Mainz.

\section{Searches for Solar Axions}
\label{sec: solar axions}

% Head 2
\subsection{Solar Axions}
\label{sec:solar_axions}

\begin{figure}[b]
  \centering
\includegraphics[height=9cm]{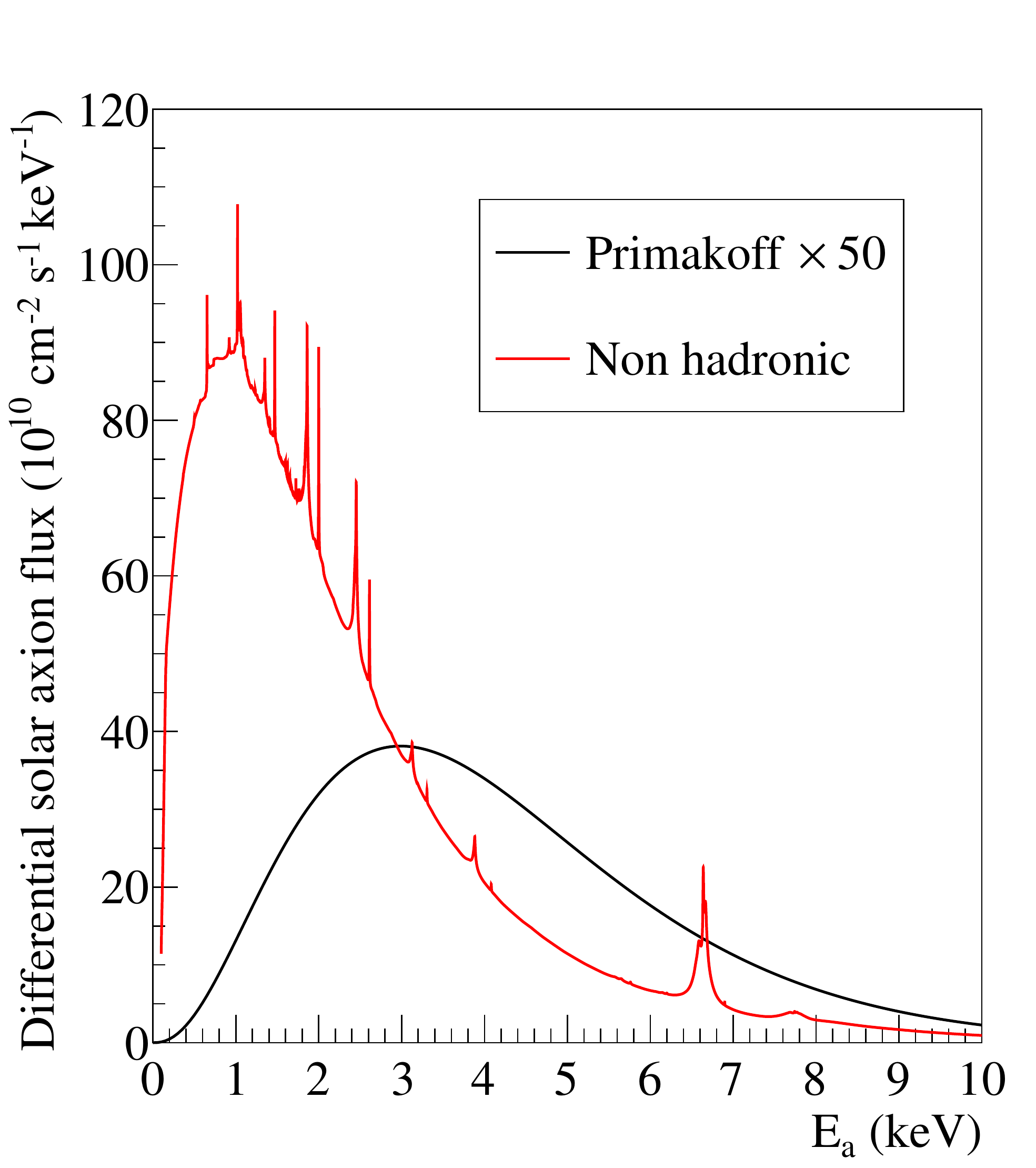}
 \caption{ Solar axion flux due to the standard Primakoff conversion (in black) for $g_{a \gamma \gamma}=10^{-12}$ GeV$^{-1}$, as well as from ABC reactions (in red) with $g_{aee}=~10^{-13}$. The Primakoff spectrum has been scaled up by a factor 50 to make both contributions comparable in the plot. }
\label{fig:axion_flux}
\end{figure}

Axions can be produced in the solar interior by the Primakoff conversion of plasma photons into axions in the Coulomb field of charged particles via the generic $a\gamma\gamma$ vertex~\cite{Raffelt:1999tx}, giving rise to a solar axion flux at the Earth surface~\cite{Andriamonje:2007ew} of $\Phi_{\rm a}=g_{10}^2\,  3.75\times10^{11}~{\rm cm}^{-2}~{\rm s}^{-1}$ (where $g_{10} = g_{a \gamma \gamma} / 10^{-10}$~GeV$^{-1}$), which corresponds to a fraction of the solar luminosity of $\mathcal{L}_{\rm a}/\mathcal{L}_\odot = g^2_{10} 1.85\times 10^{-3} $. These axions have a broad spectral distribution around 1$-$10~keV, determined by the solar core's temperature, and usefully parameterized by the following expression~\cite{Andriamonje:2007ew}:

\begin{eqnarray}\label{solar_spectrum}
\frac{{\rm d}\Phi_{\rm a}}{{\rm d}E}=6.02\times
  10^{10}~{\rm cm}^{-2}~{\rm s}^{-1}~{\rm keV}^{-1}\, g_{10}^2
  \,E^{2.481}{\rm e}^{-E/1.205}\,\,\,\,(E~{\rm in~keV})
\end{eqnarray}

\noindent that is plotted in Fig.~\ref{fig:axion_flux}. This is a robust prediction involving well-known solar physics and the generic $a\gamma\gamma$ vertex (and thus also valid for more generic ALPs). For particular non-hadronic axion models having a tree-level coupling with electrons $g_{aee}$, other productions channels like axion recombination, bremsstrahlung or Compton (i.e. ABC reactions~\cite{Redondo:2013wwa}) should be taken into account. As shown in Fig.~\ref{fig:axion_flux}, if present, this additional solar axion flux could be a factor $\sim 10^2$ larger that the standard Primakoff one, while having lower energies, peaking at $\sim$1 keV. However, astrophysical limits on $g_{aee}$ are quite restrictive and largely disfavour the values that could be reached by helioscopes looking at the non-hadronic solar axion flux.  In the future IAXO (see Section \ref{Sec: IAXO}) may have sensitivity to ABC solar axions for non-excluded values of $g_{aee}$.
Finally, axion emission in solar nuclear transitions, by virtue of axion-nucleon interactions has also been considered in the literature.

By means of the $a\gamma\gamma$ vertex, solar axions can be efficiently converted back into photons in the presence of an electromagnetic
field. In crystalline detectors~\cite{Buchmuller:1989rb,Paschos:1993yf,Creswick:1997pg}, this effect gives rise to  characteristic Bragg patterns that have been searched for as by-products of a number of underground WIMP experiments~\cite{Avignone:1997th,Morales:2001we,Bernabei:2001ny,Ahmed:2009ht,Armengaud:2013rta}. However, the prospects of this technique have proven limited~\cite{Cebrian:1998mu,Avignone:2010zn} and do not compete with dedicated helioscope experiments that use a powerful magnet to effect the conversion. Solar axion detection by means of the axioelectric effect in the detector~\cite{Ljubicic:2004gt,Derbin:2011gg,Derbin:2011zz,Derbin:2012yk,Bellini:2012kz} or, for monochromatic axions emitted in solar nuclear transitions, the resonant absorption by the same nuclide at the detector~\cite{Moriyama:1995bz,Krcmar:1998xn,Krcmar:2001si,Derbin:2009jw}, have also been considered. While interesting for some specific WISP models, all these techniques remain far from the sensitivity required to probe QCD axion models and/or parameter space not excluded by astrophysics. So far only axion helioscopes have reached relevant QCD axion parameter space.

\subsection{Axion helioscopes}
\label{sec:helioscopes}

The probability that an axion going through the transverse magnetic field $B$ over a
length $L$ will convert to a photon is given by \cite{Sikivie:1983ip,Zioutas:2004hi,Andriamonje:2007ew}:

\begin{eqnarray}\label{conversion_prob}
  P_{a\gamma} = 2.6 \times 10^{-17} \left(\frac{B}{10 \mathrm{\ T}}\right)^2
  \left(\frac{L}{10 \mathrm{\ m}}\right)^2 \nonumber \left(g_{a \gamma \gamma} \times 10^{10}
  \mathrm{\ GeV}\right)^2 F
\end{eqnarray}

\noindent where the form factor $F$ accounts for the coherence of the conversion process:

\begin{equation}\label{matrix_element}
    F=\frac{2(1-\cos q L)}{(qL)^2}
\end{equation}

\noindent and $q$ is the momentum transfer. The fact that the axion
is not massless implies that the axion and photon states will gradually slip out of phase with distance. The coherence is preserved
($F \simeq 1$) as long as $qL \ll 1$,
which for solar axion energies and a magnet length of $\sim$10 m
happens for axion masses up to $\sim
10^{-2}$ eV, while for higher masses $F$
begins to decrease, and so does the sensitivity of the experiment.
To mitigate the loss of coherence, a buffer gas can be introduced into the magnet
beam pipes \cite{vanBibber:1988ge,Arik:2008mq} to impart an effective
mass to the photons $m_\gamma = \omega_{\rm p}$
(where $\omega_{\rm p}$ is the plasma frequency of the gas,
 $\omega_{\rm p}^2=4\pi\alpha n_e/m_e$).
 %, being $n_e$ the density of electrons and $m_e$ the electron mass)
% $\omega_p^2=4\pi n_e r_0$,  $n_e$ is the density of
%electrons and $r_0$ is the classical electron radius)
For axion masses that match the photon mass, $q=0$ and full
coherence is restored. By changing the pressure of the gas inside
the pipe in a controlled manner, the photon mass can be
systematically increased and the sensitivity of the experiment can
be extended to higher axion masses.

% Margin note
\begin{marginnote}
\entry{IAXO}{International AXion Observatory}
\entry{CAST}{CERN Axion Solar Telescope}
\end{marginnote}

The basic layout of an axion helioscope thus requires a powerful magnet
coupled to one or more x-ray detectors. When the magnet is aligned
with the Sun, an excess of x-rays at the exit of the magnet is
expected, over the background measured at non-alignment periods. This detection concept was first experimentally realized
at Brookhaven National Laboratory (BNL) in 1992. A stationary dipole magnet with a field of $B = 2.2$~T and a length of $L = 1.8$~m was oriented towards the setting Sun~\cite{Lazarus:1992ry}. The experiment derived an upper limit on $g_{a \gamma \gamma}$
$(99\%$ CL) $< 3.6\times10^{-9}$ GeV$^{-1}$ for $m_a < 0.03$ eV. At the University of Tokyo, a second-generation experiment was built: the SUMICO axion heliscope. Not only did this experiment implement dynamic tracking of the Sun but it also used a more powerful magnet ($B =$ 4 T, $L =$ 2.3 m) than its BNL predecessor. The bore, located between the two coils of the magnet, was evacuated and higher-performance detectors were installed~\cite{Inoue:2002qy,Moriyama:1998kd,Inoue:2008zp}. This new setup resulted in an improved upper limit in the mass range up to 0.03 eV of $g_{a \gamma \gamma} (95\%~{\rm CL} ) < 6.0\times 10^{-10}$ GeV$^{-1}$. Later experimental improvements included the additional use of a buffer gas to enhance sensitivity to higher-mass axions.

\subsection{CAST}
\label{sec:cast}

\begin{figure}[b] \centering
%\includegraphics[width=10cm]{IAXO_Zoom.pdf}
%\includegraphics[height=7.5cm]{scenarios.eps}
%\includegraphics[height=7.5cm]{ExclusionPlotCaCe.eps}\hspace{2pc}%
%\caption{\label{fig:closeup} Close-up of the high mass part of parameter space of Fig.~\ref{fig:paraspace} (1 meV $<m_a<$ 1 eV). The two lines correspond to two different set of assumptions (more or less conservative) to compute IAXO sensitivity. Plot from ~\cite{Irastorza:1567109}.}
%\includegraphics[width=10cm]{ALPspace.pdf}
\includegraphics[height=10cm]{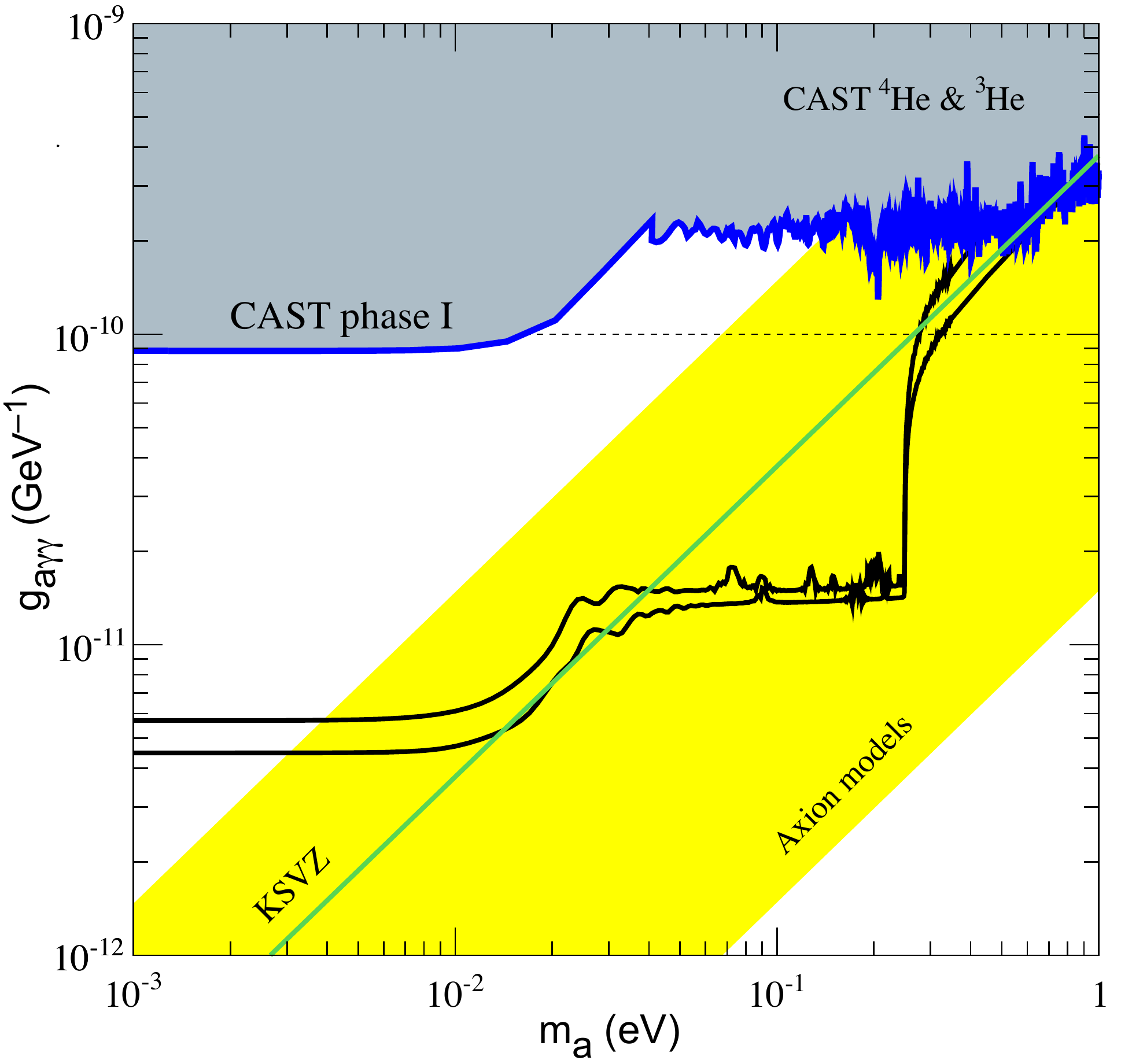}
\caption{\label{fig:paraspace} ALP parameter space ($\gagamma - m_a$) with the region (yellow band) where QCD axion models lie. Solid areas indicate the regions excluded by current experiments, among them CAST, while the two black lines indicate the expected sensitivity of the future IAXO under two different sets of assumptions.}
\end{figure}

A third-generation experiment, the CERN Axion Solar Telescope (CAST), began data collection in 2003 and is still in operation. The experiment uses a Large Hadron Collider (LHC) dipole prototype magnet with a magnetic field of up to 9 T over a length of 9.3 m~\cite{Zioutas:1998cc}. CAST is able to follow the Sun for several hours per day using an elevation and azimuth drive. This CERN experiment is the first helioscope to employ x-ray focusing optics for one of its four detector lines~\cite{Kuster:2007ue}, as well as low background techniques from detectors in underground laboratories~\cite{Abbon:2007ug,Aune:2013pna,Aune:2013nza}. During its observational program from 2003 to 2011, CAST operated first with its magnet bores under vacuum (2003--2004) to probe masses $m_a < 0.02$~eV. No significant signal above background was observed. Thus, an upper limit on the axion-to-photon coupling of $g_{a \gamma \gamma}~(95\%~$CL$) < 8.8\times 10^{-11}$ GeV$^{-1}$ was obtained~\cite{Zioutas:2004hi,Andriamonje:2007ew}. The experiment was then upgraded to be operated with $^4$He (2005--2006) and
$^3$He gas (2008--2011) to obtain continuous, high sensitivity up to an axion mass of $m_a = 1.17$ eV. Data released up to now provide an average limit of $g_{a \gamma \gamma}~(95\%~$CL$) \lesssim 2.3\times10^{-10}$ GeV$^{-1}$, for the higher mass range of 0.02 eV $< m_a <$ 0.64 eV~\cite{Arik:2008mq,Aune:2011rx} and of about $g_{a \gamma \gamma}~(95\%~$CL$) \lesssim 3.3\times10^{-10}$ GeV$^{-1}$ for 0.64 eV $< m_a <$ 1.17 eV~\cite{Arik:2013nya}, with the exact value depending on the pressure setting. The envelope of all these limits is shown in Fig.~\ref{fig:paraspace}. Byproducts of CAST include the search for 14.4 keV solar axions emitted in the M1 transition of $^{57}$Fe nuclei~\cite{Andriamonje:2009dx}, the search from MeV axions from $^7$Li and D($p,\gamma$)$^3$He nuclear transitions~\cite{Andriamonje:2009ar}, the search for solar axions from ABC reactions~\cite{Barth:2013sma}, and the search of more exotic ALP or WISP models, like chameleon particles also potentially emitted in the sun~\cite{Brax:2011wp,Baum:2014rka}.
As part of the R\&D to assess the technologies for the next generation axion helioscope IAXO, new lower background Micromegas detectors are actively being developed~\cite{Aune:2013pna,Aune:2013nza}, and a new x-ray telescope coupled with one such detector being built and installed in CAST in 2014. This improved equipment is currently allowing CAST to revisit the $^4$He (in 2012) and vacuum (in 2013-15) configurations with an incremental improvement in sensitivity.

So far each subsequent generation of axion helioscopes has resulted in an improvement in sensitivity to the axion-photon coupling constant of about a factor 6 over its predecessors. CAST has been the first axion helioscope to surpass the stringent limits from astrophysics $g_{a \gamma \gamma} \lesssim 10^{-10}$ GeV$^{-1}$ over a large mass range and to probe allowed ALP parameter space. As shown in Fig.~\ref{fig:paraspace}, in the region of higher axion masses ($m_a \gtrsim 0.1$ eV), the experiment has entered the band of QCD axion models for the first time and excluded KSVZ axions of specific mass values. CAST is the largest collaboration in axion physics with $\sim$ 70 physicists from about 16 different institutions in Europe and the USA. The scalability of the helioscope technique has been recently proven~\cite{Irastorza:2011gs} and a substantial step beyond CAST state-of-the-art is achievable by the proposed International Axion Observatory (IAXO).

\subsection{IAXO}
\label{Sec: IAXO}

\begin{figure}[t] \centering
%\psfrag{Shielding}{Shielding}
%\includegraphics[width=5cm]{magnet1.eps}
%\includegraphics[width=11cm]{sketch_helioscope.pdf}\hspace{2pc}%
\includegraphics[width=\columnwidth]{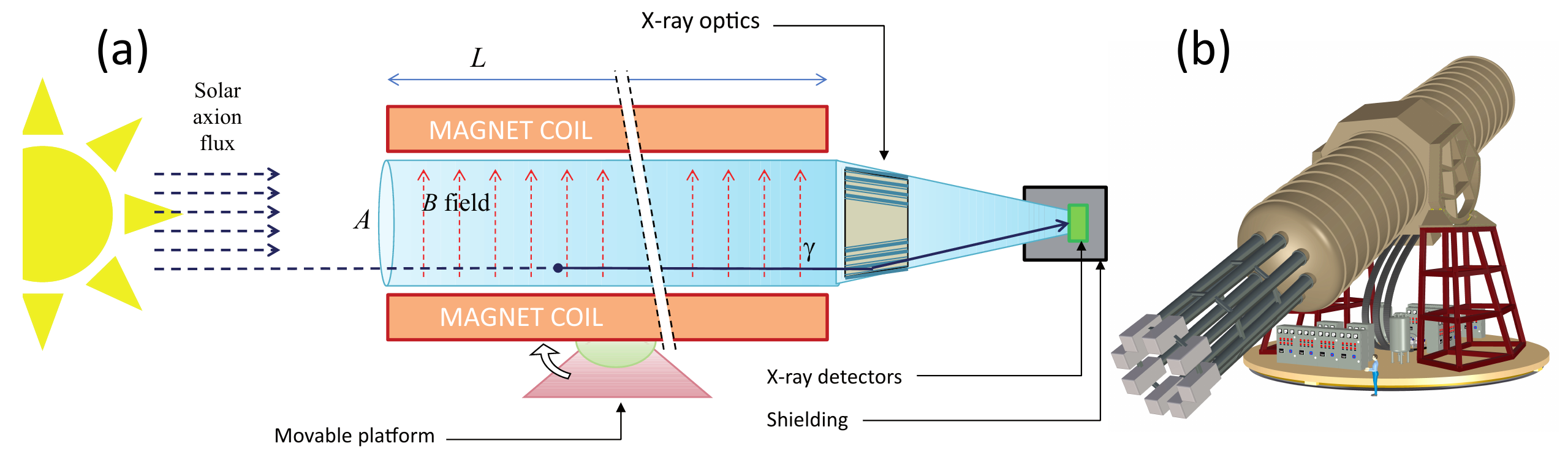}\hspace{2pc}%
\caption{\label{fig:NGAH_sketch} (a) Conceptual arrangement
of an enhanced axion helioscope with x-ray focusing. Solar axions are converted into photons by the transverse magnetic field inside the bore of a powerful magnet. The resulting quasi-parallel beam of photons of cross sectional area $A$ is concentrated by  appropriate x-ray optics onto a small spot area $a$ in a low background detector.
(b) The envisaged design for IAXO includes eight such magnet bores, with their respective optics and detectors. }
\end{figure}

%\begin{figure}[b]
%\centering
%\includegraphics[height=6cm] {system_wo_dome.jpg}
% \caption{ Conceptual design of IAXO~\cite{Armengaud:2014gea}.}
%\label{fig:iaxo_sketch}
%\end{figure}

Going substantially beyond CAST sensitivity is possible only by going to a new magnet, designed and built maximizing the helioscope magnet's figure of merit $f_M  = B^2\:L^2\:A$, proportional to the photon signal from converted axions, where $B$, $L$ and $A$ are the magnet's field strength, length and cross sectional area, respectively \cite{Irastorza:2011gs}.
%Improving CAST $f_M$ can only be achieved~\cite{Irastorza:2011gs} by a different magnet configuration with a much larger magnet aperture $A$, which in the case of the CAST magnet is only $3\times10^{-3}$~m$^2$.
However, for this figure of merit to directly translate into signal-to-noise ratio of the overall experiment for a large aperture magnet, the entire cross sectional area of the magnet must be equipped with x-ray focusing optics. The layout of this \textit{enhanced axion helioscope}, sketched in Figure \ref{fig:NGAH_sketch}, was proposed~\cite{Irastorza:2011gs} as the basis for IAXO. The project is at the point of conceptual design, recently finished~\cite{Armengaud:2014gea}, and it is now moving into the realization of the technical design report, including some prototyping activities. A recent Letter of Intent~\cite{Irastorza:1567109} to CERN was positively reviewed.

%Contrary to previous helioscopes, IAXO's magnet will follow a toroidal configuration~\cite{Shilon:2012te}
To produce an intense magnetic field over a large volume, and maximize the figure of merit within realistic limits of the different technologies in play, motivates moving to a 25 m long and 5.2 m diameter toroid assembled from 8 coils, producing 2.5 tesla in 8 bores of 600 mm diameter \cite{Shilon:2012te}. The magnet is supported by a tracking system similar to that of large telescopes. Figure~\ref{fig:NGAH_sketch} shows the conceptual design of the infrastructure~\cite{Armengaud:2014gea}. Each of the bores will be equipped with x-ray optics similar to those used on NASA's NuSTAR~\cite{nustar2013}, an x-ray astrophysics satellite with two focusing telescopes that operate in the 3 - 79 keV band, consisting of thousands of thermally-formed glass substrates deposited with multilayer coatings. For IAXO, the multilayer coatings will be designed to match the solar axion spectrum~\cite{doi:10.1117/12.2024476}. At the focal plane in each of the optics, IAXO will have small gaseous chambers read by pixelated planes of Micromegas,  surrounded by active and passive shielding. These detectors are being  developed~\cite{Irastorza2011_EAS,Dafni:2012fi,Dafni:2012zz,trexwebpage} for rare event searches and they show promise to reach background levels below $\sim$$10^{-7}$~\ckcs in IAXO~\cite{Aune:2013pna,Aune:2013nza}. These levels are achieved by the use of radiopure detector components,  shielding, and offline discrimination-algorithms on the 3D event topology in the gas registered by the pixelised readout.

IAXO will have 5 orders of magnitude better signal-to-noise ratio than CAST, which means a sensitivity to $\gagamma$ values as low as, or even surpassing, $\gagamma \sim 5 \times 10^{-12}~{\rm GeV}^{-1}$, for a wide range of axion masses up to about 0.01~eV after $\sim3$ years of data taking in vacuum, and around $\gagamma \sim$10$^{-11}~{\rm GeV}^{-1}$ up to about 0.25 eV, after an additional $\sim3$ years with the use of a buffer gas in the conversion bores. IAXO will thus deeply enter into unexplored ALP and axion parameter space, as indicated by Fig.~\ref{fig:paraspace}. At the lowest mass values, IAXO will test ALP models  invoked to explain the anomalies in light propagation over astronomical distances. At the high mass part, $m_a > 1 $ meV, IAXO will explore a broad range of QCD axion models. Its sensitivity would reach axion models with masses down to the few meV range, superseding the SN~1987A energy loss limits on the axion mass and entering parameter space of progressively higher cosmological interest. At the higher part of the range (0.1 - 1 eV) axions are good candidates to the hot DM or additional \textit{dark radiation} that could restore agreement in cosmological parameters. IAXO could also be sensitive to ABC solar axion for non-excluded values of $g_{aee}$ and thus directly test the models invoked to solve the anomalous cooling observed for white dwarfs~\cite{Isern:2008fs,Corsico:2012ki,Corsico:2012sh}. Additional equipment beyond the baseline configuration (like InGrid detectors, Transition Edge Sensors or low noise CCDs~\cite{Irastorza:1567109}) would  extend the detection energy window, and thus explore other less standard physics cases (e.g. the possibility to directly detect the cosmic axion background predicted by some dark radiation models). Finally, the possibility to equip the huge magnetic volume of IAXO with microwave cavities or antennas sensitive to relic axions is being studied, with promising preliminary projections~\cite{redondo_patras_2014} in mass ranges complementary with previous haloscope searches.

%\begin{figure}[t]
%\centering
%\includegraphics[height=7cm]{IAXO_Zoom.pdf}
% \caption{ Close-up of the high mass part of parameter space of Fig.~\ref{fig:paraspace} (1 meV $<m_a<$ 1 eV). The two lines correspond to two different set of assumptions (more or less conservative) to compute IAXO sensitivity. Plot from ~\cite{Irastorza:1567109}.}
%\label{fig:highmass}
%\end{figure}

\section{Purely Laboratory Experiments}

%Purely laboratory based light-shining-through-a-wall (LSW) experiments (see Figure~\ref{fig:regcav}) offer full control on the production of WISPs and the re-generation of photons, thus depending hardly on theoretical uncertainties of WISP physics. 
%
%
%In the following we discuss the status and prospects of LSW experiments searching for axion-like particles.
%It should be noted however that LSW experiments also allow to search for other WISPs like hidden photons or mini-charged particles~\cite{Dobrich:2012sw,Dobrich:2012jd}.

\begin{figure}[htb]
\includegraphics[width=120mm]{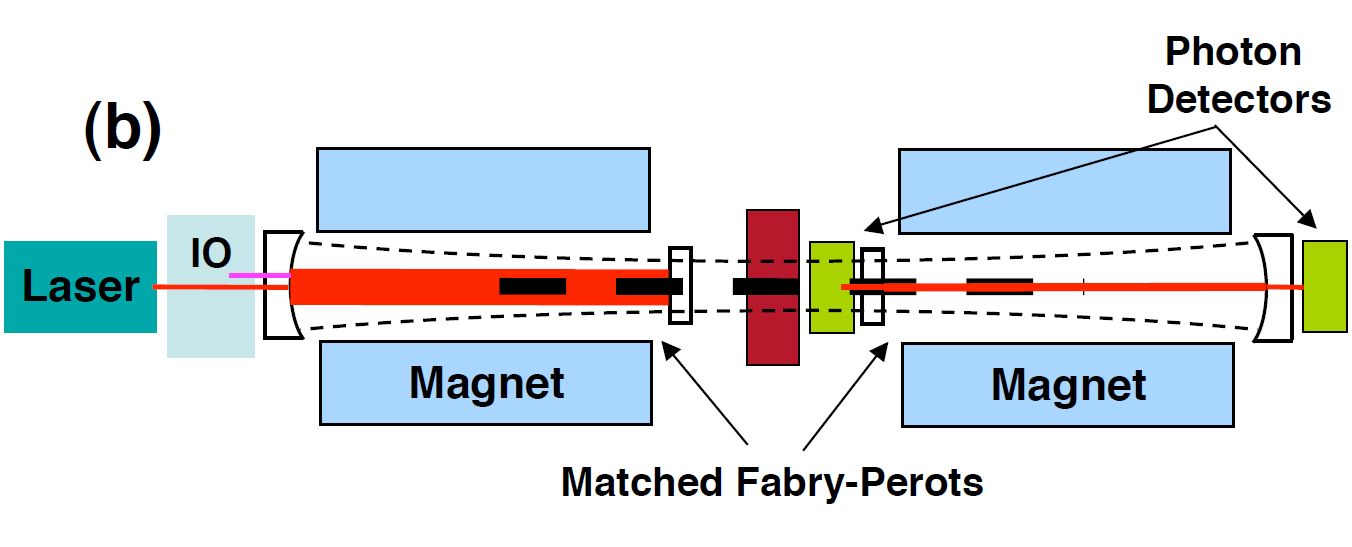}
\caption{
The principle of photon regeneration.  Current and future experiments will employ Fabry-Perot cavities both for the production and regeneration regions, and actively locked together, to greatly enhance the experimental sensitivity.  The ALPS I experiment has already used such a cavity in the production region \cite{Ehret:2010mh,Ehret:2009sq}.}
%The principle of a light-shining-through-a-wall (LSW) experiment and the optical concept of future LSW experiments:
%light is shone into a magnetic field. ALPs could be created by interaction of the laser light with the magnetic field. A barrier blocks the light, but WISPs easily traverse it. In the second part of the experiment behind the wall some WISPs convert back to photons giving rise to the impression of light shining through a wall.
%With two matched Fabry-Perots cavities the sensitivity will be greatly enhanced by ``recycling" laser light before the wall and by boosting the back-conversion probability of WISPs into light behind the wall. A setup with the left cavity has been realized already at the ALPS~I experiment~\cite{Ehret:2010mh,Ehret:2009sq}. The figure was taken from \cite{Mueller:2009wt}. PLEASE CHECK FOR COPYRIGHT!}
 \label{fig:regcav} 
\end{figure}

Recent years have seen burgeoning interest in experiments searching for axions or other light exotica which do not rely on cosmological or astrophysical sources.  This section concerns photon regeneration, colloquially referred to as ``Light Shining through Walls" (LSW) by which photons mix with axions (or WISPs more generally), the axions (or WISPs) being regenerated into the photon state on the other side of an optical barrier (Figure \ref{fig:regcav}).  For axions, production and regeneration require a transverse magnetic field, but magnets may or may not be required for WISPs in general.  We do not attempt to cover the sector of short-range spin dependent forces, which would properly require a review of its own.

The most sensitive LSW experiments today use or plan to use coherent light to provide the photons before the wall. Such light offers the possibility to enhance sensitivities by implementing Fabry-Perots cavities (Figure~\ref{fig:regcav}):
\begin{itemize}
\item A cavity in front of the wall enables recycling the light shone against the wall and hence to increase the effective light power by a power-built-up factor $\mathcal{F}_{PC}$.
\item A cavity with a power-built-up factor $\mathcal{F}_{RC}$ behind the wall increases the reconversion probability.
%\footnote{The physics of the resonator behind the wall is very similar to the concept of cavity microwave experiments searching for dark matter WISPs.} of a WISP with a factor $Q_{RC}$. 
This technique is called ``resonant regeneration"\footnote{The factor $\mathcal{F}$ is in fact the Finesse of the Fabry-Perot resonator.  The quality factor, $Q$ of the resonator, which determines the intrinsic line-width, is given by $Q = \left( l/\lambda \right)  \mathcal{F}$, where $l$ is the length of the cavity, and $\lambda$ the wavelength.}.
\end{itemize}
The influence of the empty cavity behind the wall is  similar to the Purcell effect described first in 1946~\cite{Purcell:1946} (see also ~\cite{Haroche:1989,Haroche:1991kj} for related experiments).
The implementation of resonating cavities in LSW experiments has been worked out in more detail in~\cite{Hoogeveen:1990vq,Fukuda:1996kwa,Mueller:2009wt}.
As $\mathcal{F}$ factors of several $10^4$ in the optical regime and even several $10^5$ for microwaves can be reached, resonators allow for significant improvements of experimental sensitivities.  

\subsection{LSW experiments with microwaves, optical photons and X-rays}
\label{sec:LSWexperiments}
Experiments using microwaves usually work in the near-field approximation. One can picture this situation as having a production cavity (PC) emitting a beam of WISPs, where the shape is given by the mode resonating in the cavity. Hence one has to place the detector (regeneration) cavity (RC) somewhere next to the production cavity so that WISPs pass through it. The probability for a photon-WISP-photon oscillation ($\gamma \rightarrow \phi \rightarrow \gamma$) with an effective photon-WISP coupling $C_{wisp}$ in such an installation~\cite{Jaeckel:2007ch,Caspers:2009cj} is given by Eq.~\ref{eq:LSWmicro}, where the geometrical form factor G describes the overlap of the modes resonating in the PC and RC. The size of such experiments is essentially given by the microwave wavelength.
\begin{equation}
P_{\gamma \rightarrow \phi \rightarrow \gamma} = |C_{wisp}|^4 \cdot |G|^2 \cdot \mathcal{F}_{PC} \cdot \mathcal{F}_{RC}
\label{eq:LSWmicro}
\end{equation}
 
In a far-field approximation (as realized with optical photons and X-rays) the conversion probability is given by Eq.~\ref{eq:LSWoptic}.
\begin{equation}
P_{\gamma \rightarrow \phi \rightarrow \gamma}= 
\frac{\omega}{\sqrt{\omega^2-m_\phi^2}} \cdot |C_{wisp}|^4 \cdot \mathcal{F}_{PC} \cdot \mathcal{F}_{RC} \cdot \sin^4\left(\frac{q \cdot l}{2}\right)
\label{eq:LSWoptic}
\end{equation}

Here $ q =| n\cdot \omega - \sqrt{\omega^2-m_\phi^2} | $ with the photon energy $\omega$, the WISP mass $m_\phi$ and the refractive index $n$, and $l$ the length of the experiment in front of and behind the wall.
For hidden photons and axion-like particles, the couplings are given by Eq.~\ref{eq:LSWcoupling}:
\begin{equation}
|C_{hp}|^2 = 4 \chi^2\cdot\frac{m_\phi^4}{\left(m_\phi^2+2\omega^2(n-1)\right)^2};\ 
|C_{alp}|^2 = 4\cdot \frac{(g_{a\gamma\gamma} \omega B)^2}{\left(m_\phi^2+2\omega^2(n-1)\right)^2}
\label{eq:LSWcoupling}
\end{equation}
with the dimensionless parameter $\chi$ and the two-photon coupling strength $g_{a\gamma\gamma}$ with the dimension mass$^{-1}$. In Eq.~\ref{eq:LSWcoupling} it is assumed that the homogeneous magnetic field is oriented perpendicular (parallel) to the E-field of the light wave for interacting with scalar (pseudoscalar) ALPs.
The maximal sensitivity is obtained in vacuum with $n=1$.
Following the equations \ref{eq:LSWoptic} and \ref{eq:LSWcoupling} one arrives at the approximation in equation\,\ref{eq:ALPcoupling} for $ql\ll 1$ and $n=1$:
\begin{equation}
P_{\gamma \rightarrow \phi \rightarrow \gamma} =
\frac{1}{16} \cdot \mathcal{F}_{PC} \mathcal{F}_{RC} \cdot \left( g_{a\gamma\gamma}Bl \right)^4 = 
6\cdot 10^{-38}\cdot \mathcal{F}_{PC} \mathcal{F}_{RC} \cdot \left(\frac{g_{a\gamma\gamma}}{10^{-10}GeV^{-1}}\frac{B}{1\,T}\frac{l}{10m} \right)^4 
\label{eq:ALPcoupling}
\end{equation}
%The challenges for experiments to detect such low probability processes is obvious.
The leading microwave LSW experiment is the CERN Resonant Weakly Interacting sub-eV Particle Search, CROWS \cite{Betz:2013dza}.
Its has fully exploited the resonant regeneration technique described above and achieved about the same sensitivity as the optical experiments ALPS~I  at DESY~\cite{Ehret:2010mh} and OSQAR at CERN~\cite{Ballou:2014myz} which had no cavity behind the wall.
ALP-photon couplings around $g_{a\gamma\gamma}=5\cdot 10^{-8} GeV^{-1}$ have been probed.
While CROWS has searched for ALPs up to its kinematic limit $m_\phi = \omega$, the optical experiments were limited to roughly $ m_\phi < 10^{-3} \omega$ due to Eq.~\ref{eq:LSWoptic}. 

LSW experiments have also been performed  with intense X-ray beams available at synchrotron radiation sources\,\cite{Battesti:2010dm,Inada:2013tx}. 
Due to the higher photon energy such experiments can probe WISPs up to eV-masses and beyond, but due to the relative low photon number flux and the present impossibility to implement cavities with high power built-ups, X-ray based experiments do not reach the sensitivity of optical or microwave LSW. 

\subsection{The future of LSW experiments}
\label{sec:LSWfuture}
\begin{figure}[htb]
\includegraphics[width=120mm]{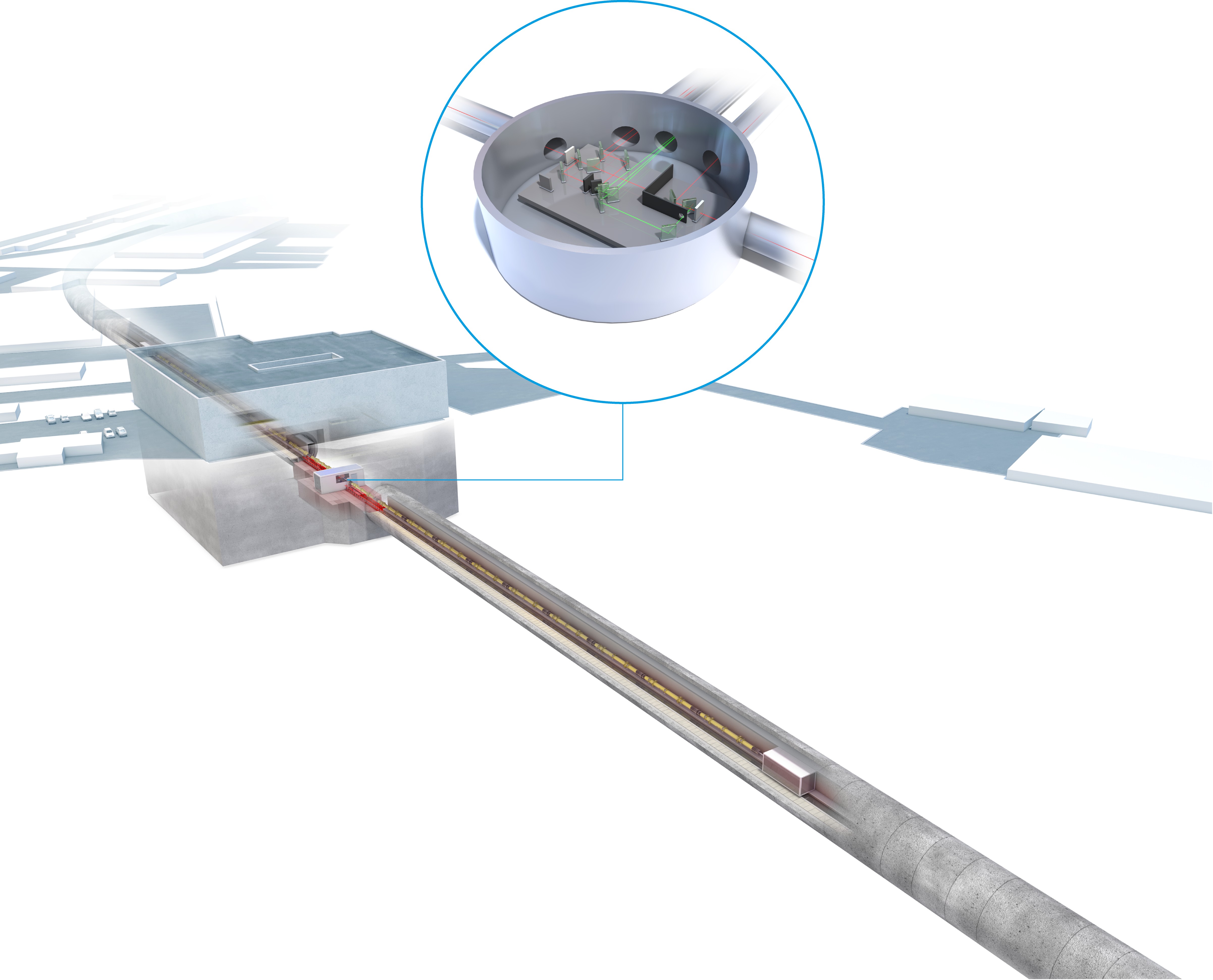}
\caption{An artist's view of the ALPS~II experiment under preparation at DESY:
it is planned to install a string of 20 HERA dipole magnets  in a straight section of the tunnel of the decommissioned HERA accelerator. The insert sketches parts of the central optics used to align and control the two optical resonators in front of and behind the wall. Copyright 2013, DESY.}
 \label{fig:alpsII} 
\end{figure}

%New ideas to improve  LSW experiments searching for ALPs focus on the optical or infrared regime as the length of such installations can be increased considerably and an implementation of regeneration cavities promise a jump in sensitivity (equation\,\ref{eq:ALPcoupling}).
The most advanced proposal is the ALPS~II~\cite{Bahre:2013ywa} project being under preparation at DESY (see Figure~\ref{fig:alpsII}).
In Table\,\ref{tab:param} critical parameters of ALPS~I, ALPS~II and a future experiment JURA (Joint Undertaking on Research for Axion-like particles) are listed. 
%JURA serves for illustration purposes only; currently there is no concrete project behind this idea.
%
\begin{table}
\begin{center}
\begin{tabular}{|l|c|c|c|c|c|}
\hline
Parameter & Sensitivity & ALPS~I & ALPS~II & JURA \\[1pt] \hline
Effective laser power $P_{laser}$ & $g_{a\gamma\gamma} \propto P_{laser}^{-1/4}$ & 1\,kW & 150\,kW & 1000\,kW\\
$\mathcal{F}_{RC}$ & $g_{a\gamma\gamma} \propto \mathcal{F}_{RC}^{-1/4}$ & 1 & 40,000 & 100,000 \\[1pt] \hline
Length (B field) $l$ & $g_{a\gamma\gamma} \propto (l)^{-1}$ & 4.4\,m & 88\,m & 286\,m\\
Magnetic field $B$ & $g_{a\gamma\gamma} \propto (B)^{-1}$ & 5.0\,T & 5.3\,T & 13\,T\\ [1pt]  \hline
\end{tabular}
\end{center}
\caption{\label{tab:param} Parameters of the ALPS~I~\cite{Ehret:2010mh} experiment in comparison to ALPS~II~\cite{Bahre:2013ywa} and a hypothetical future experiment JURA, just in conceptual phase. The second column shows the dependence of the reachable ALP-photon coupling on the experimental parameters.}
\end{table}
It is evident that ALPS~II has the potential to increase the sensitivity for $g_{a\gamma\gamma}$ by more than three orders of magnitude reaching down to $g_{a\gamma\gamma} = 2 \times 10^{-11} GeV^{-1}$ thus going beyond present-day limits from astrophysics.
%In the following we will briefly sketch the concepts for optics, detectors and magnets under development for ALPS~II and possible future experiments.\\

To realize an experiment as shown in Figures\,\ref{fig:regcav} and \ref{fig:alpsII} both cavities must be mode matched and phase locked. 
%This can be pictured as being able (theoretically) to remove the two central flat mirrors and the wall while the long cavity (now defined by the two outer curved mirrors) stays resonating.
At ALPS~II continuous wave infrared light (1064\,nm) with a power of up to 35\,W is provided by a single-mode single-frequency laser system. 
Through a curved mirror the laser light is injected into the cavity in front of the wall which is locked via the Pound-Drever-Hall (PDH) sensing scheme~\cite{Black:2001}.
This ALPS~II cavity is designed for a power build-up $\mathcal{F}_{RC} = 5,000$ providing a circulating power of 150\,kW and limiting the power density on the mirrors to 500\,kW/cm$^2$, about an order of magnitude below their damage thresholds.
To allow for locking the cavity behind the wall with the PDH method, a fraction of the 1064\,nm light is frequency-doubled to $532\,$nm in a KTP crystal and fed into the cavity targeting for a power build-up of $\mathcal{F}_{RC} = 40,000$. 
%Each WISP reconverting behind the wall will create one ${\rm 1064\,nm}$ photon, so one has to exclude any background in this wavelength region (as induced by $532\,$nm photons for example).
For future experiments beyond ALPS~II one could think of increasing the circulating light power in front of the wall by another order of magnitude (see also~\cite{Mueller:2009wt}) and to improve on $\mathcal{F}_{RC}$ by advanced seismic isolations of the optical components for example.\\

ALPS~II will use a Transition Edge Sensor (TES)~\cite{Lita:2010,Miller:2011} based on a thin superconducting Tungsten film (${\rm 25\,\mu m \cdot 25\,\mu m \cdot 20\, nm}$) as photon detectors.
1064\,nm photons appear as fast pulses with a decay time of about $1.5 \, \mu s$.
The intrinsic background is below $10^{-4}$ counts per second, the energy resolution for single infrared photons is better than $10\%$ and the whole system can be operated stably over long time scales~\cite{Dreyling-Eschweiler:2014eya, Dreyling-Eschweiler:2014mxa,Dreyling-Eschweiler:2014MO}. 
Infrared photon detection sensitivity well below mHz rates should be achievable.

An alternative detection scheme is proposed in~\cite{Mueller:2009wt}.
Here the regeneration cavity is controlled by light which frequency is only shifted by a few free spectral ranges compares to the light in the production cavity (not frequency-doubled as for ALPS~II). 
This approach would allow for a heterodyne detection of the reconverted photons behind the wall by mixing their signal (which would occur at the production cavity's light frequency) with the light controlling the regeneration cavity. 
In principle any detector background noise can be rendered negligible in this scheme.\\

The coupling $g_{a\gamma\gamma}$ that can be reached goes as $(Bl)^{-1}$, but at present LSW experiments have used one or two dipole magnets only.
The maximal length of an LSW experiment depends on the aperture of the magnets and the laser beam divergence. 
%If one exploits optical cavities as described above any clipping losses of the laser light have to be smaller than about $10^{-5}$.
This limits the ALPS~II setup to a total length of about 200\,m for example, which represents 20 HERA dipole magnets\footnote{The dipole magnets will be straightened to provide a sufficient aperture.}.
In the far future one could imagine magnet strings based on the dipoles under development at CERN~\cite{Bottura:2012uka,Todesco:2014gna} offering a field of about 13\,T with an aperture of about 100\,mm if the high $\rm T_c$ superconducting inner part is removed (see JURA\footnote{Using 532\,nm light instead of 1064\,nm would allow a further doubling of the length.} in Table~\ref{tab:param}).\\

The next generation of laser-based LSW experiments like ALPS~II will surpass present day limits on axion-like particles from astrophysics observation,
%They will do so with significant less theoretical uncertainties, as they try to produce and detect WISPs in the laboratory
and have the potential to probe a large fraction of the parameter space for axion-like particles as indicated by the cooling of stars or the abnormal high transparency of the universe for TeV photons.
%On the longer term LSW experiments might surpass the ALPS~II sensitivity for ALP-photon couplings by more than an order of magnitude.
However, at present no concept exists on how to access the QCD axion with purely laboratory based experiments. 
Only once the axion mass is known from dark matter experiments for example, one could think of optimizing the sensitivity of LSW experiments accordingly~\cite{Arias:2010bh}.

\section{Summary and Conclusions}

\begin{figure}
\begin{center}
\includegraphics[width= \columnwidth]{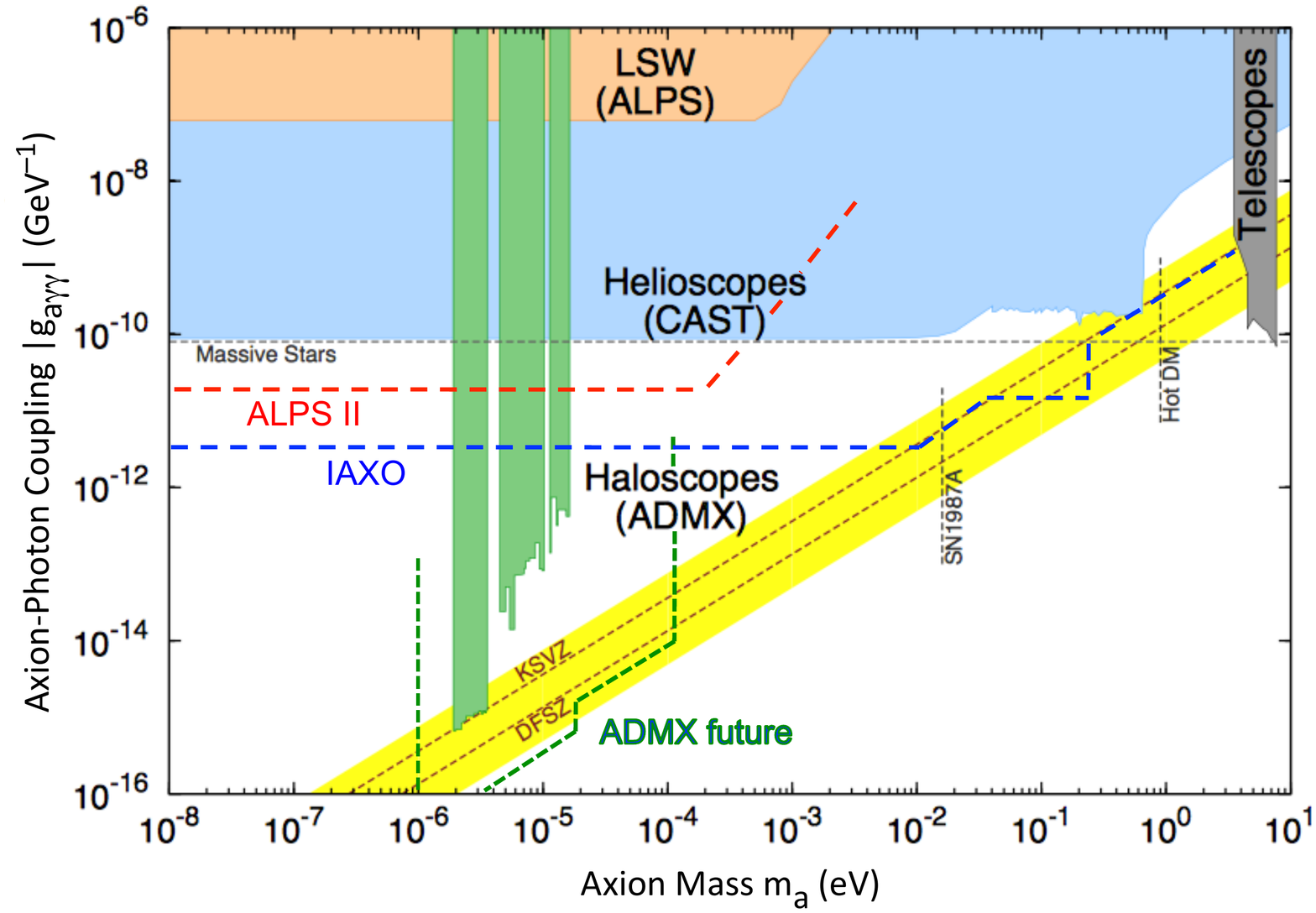}
\caption{ \label{Fig:last} Existing limits on the photon coupling of axions and axion-like particles and the projected coverage of ongoing upgrades for these experiments.  Figure adapted from Ref. \cite{Agashe:2014kda} (the Particle Data Group).}
\end{center}
\end{figure}

Figure \ref{Fig:last} schematically represents the current limits on the photon coupling of axions and axion-like particles, and the projected extension of these results with ongoing upgrades within the next five years; companion Figures \ref{Fig:CASPErWIND} and \ref{Fig:EDM} summarize the reach of future NMR experiments for lower mass axions through their nucleon coupling.  Along with experiments still in the conceptual or prototype phase, these constitute, for the first time, the emergence of a complete strategy to definitively answer the question of the existence of the axion, and its cosmological role, from the neV to meV mass range.  It is interesting to note that it has been the mass coverage rate, not the sensitivity of all dark-matter axion searches that has been the limiting factor.  If the mass were known virtually any microwave cavity experiment would be able to detect the axion, with sufficient integration time.  In fact, if the axion were discovered, it would quickly become a senior undergraduate physics laboratory experiment.  The heart of the issue with tuning high-Q experiments is that their exquisite sensitivity accrues as a trade-off for spectral bandwidth, thus necessitating tuning over a decade or more of mass. 

The searches for axion-like particles based on axion-photon mixing in the relativistic limit, i.e.~helioscope and `light shining through walls' experiments have also achieved an impressive technical sophistication and are now poised for dramatic scale-up from earlier realizations.  While their ultimate projections in $\gagamma$ might only improve on the best astrophysical limits by an order of magnitude or so, these experiments are extremely important insofar as they have discovery potential for generalized pseudoscalars that may have no origin in Peccei-Quinn symmetry, and for which the microwave cavity experiments, etc. would be irrelevant.  

The authors are hopeful, even optimistic the next major review will focus on the properties of the already-discovered axion, and the new field of axion astronomy and cosmology.

%Disclosure
%\section*{DISCLOSURE STATEMENT}
%Disclosure statement will go here.

% Acknowledgements
\section*{ACKNOWLEDGMENTS}
PWG acknowledges the support of NSF grant PHY-1316706, DOE Early Career Award DE-SC0012012,  the Terman Fellowship, and the Heising-Simons Foundation.
I.G.I acknowledges support from the Spanish Ministry of Economy and Competitiveness (MINECO) under contracts FPA2011-24058,  FPA2013-41085 and CSD2007-00042 (CPAN project), as well as from the European Research Council under the T-REX Starting Grant ERC-2009-StG-240054 of the IDEAS program of the 7th EU Framework Program.
SKL and KvB acknowledge support from the National Science Foundation, under grants PHY-1067242, and PHY-1306729, respectively.
We thank J.A. Garcia for the help with figure~\ref{fig:axion_flux}.

% References
%
% Margin notes within bibliography
%\bibnote[<vertical skip value - optional>]{Bibliography margin text.}
%
\bibliographystyle{ar-style5}
\bibliography{axionreviewbib}

\begin{thebibliography}{165}
\expandafter\ifx\csname natexlab\endcsname\relax\def\natexlab#1{#1}\fi

\bibitem{Weinberg:1977ma}
Weinberg S.
\newblock \textit{Phys. Rev. Lett.} 40:223 (1978)

\bibitem{Wilczek:1977pj}
Wilczek F.
\newblock \textit{Phys. Rev. Lett.} 40:279 (1978)

\bibitem{Peccei:1977hh}
Peccei R, Quinn HR.
\newblock \textit{Phys. Rev. Lett.} 38:1440 (1977)

\bibitem{Peccei:1977ur}
Peccei R, Quinn HR.
\newblock \textit{Phys. Rev.} D16:1791 (1977)

\bibitem{Asztalos:2006kz}
Asztalos SJ, et~al.
\newblock \textit{Ann. Rev. Nucl. Part. Sci.} 56:293 (2006)

\bibitem{Rosenberg:2000wb}
Rosenberg L, van Bibber K.
\newblock \textit{Phys. Rept.} 325:1 (2000)

\bibitem{Bradley:2003kg}
Bradley R, et~al.
\newblock \textit{Rev. Mod. Phys.} 75:777 (2003)

\bibitem{Kim:1986ax}
Kim JE.
\newblock \textit{Phys. Rept.} 150:1 (1987)

\bibitem{Cheng:1987gp}
Cheng HY.
\newblock \textit{Phys. Rept.} 158:1 (1988)

\bibitem{Turner:1989vc}
Turner MS.
\newblock \textit{Phys. Rept.} 197:67 (1990)

\bibitem{Raffelt:1990yz}
Raffelt GG.
\newblock \textit{Phys. Rept.} 198:1 (1990)

\bibitem{Kim:2008hd}
Kim JE, Carosi G.
\newblock \textit{Rev. Mod. Phys.} 82:557 (2010)

\bibitem{skl0}
Khriplovich I, Lamoreaux S.
\newblock Texts and monographs in physics. Springer-Verlag (1997)

\bibitem{skl1}
Landau L, Lifshitz E.
\newblock No. Bd. 2 in Course of theoretical physics. Butterworth Heinemann
  (1975)

\bibitem{skl2}
Baluni V.
\newblock \textit{Phys. Rev. D} 19:2227 (1979)

\bibitem{skl3}
Crewther R, Vecchia PD, Veneziano G, Witten E.
\newblock \textit{Phys. Lett. B} 88:123  (1979), erratum: 91, 487(E) (1980)

\bibitem{skl4}
Baker C, et~al.
\newblock \textit{Phys. Rev. Lett.} 97:131801 (2006)

\bibitem{skl5}
Ellis J, Gaillard MK.
\newblock \textit{Nucl. Phys. B} 150:141  (1979)

\bibitem{skl6}
Khriplovich I, Vainshtein A.
\newblock \textit{Nucl. Phys. B} 414:27  (1994)

\bibitem{Kim:1979if}
Kim JE.
\newblock \textit{Phys. Rev. Lett.} 43:103 (1979)

\bibitem{Shifman:1979if}
Shifman MA, Vainshtein A, Zakharov VI.
\newblock \textit{Nucl. Phys.} B166:493 (1980)

\bibitem{Dine:1981rt}
Dine M, Fischler W, Srednicki M.
\newblock \textit{Phys. Lett.} B104:199 (1981)

\bibitem{Zhitnitsky:1980tq}
Zhitnitsky A.
\newblock \textit{Sov. J. Nucl. Phys.} 31:260 (1980)

\bibitem{Svrcek:2006yi}
Svrcek P, Witten E.
\newblock \textit{JHEP} 0606:051 (2006)

\bibitem{Jaeckel:2010ni}
Jaeckel J, Ringwald A.
\newblock \textit{Ann. Rev. Nucl. Part. Sci.} 60:405 (2010)

\bibitem{Arias:2012az}
Arias P, et~al.
\newblock \textit{JCAP} 1206:013 (2012)

\bibitem{Ringwald:2012hr}
Ringwald A.
\newblock \textit{Phys. Dark Univ.} 1:116 (2012)

\bibitem{Meyer:2013pny}
Meyer M, Horns D, Raue M.
\newblock \textit{Phys. Rev.} D87:035027 (2013)

\bibitem{Rubtsov:2014uga}
Rubtsov G, Troitsky S.
\newblock \textit{JETP Lett.} 100:397 (2014)

\bibitem{Ayala:2014pea}
Ayala A, et~al.
\newblock \textit{Phys. Rev. Lett.} 113:191302 (2014)

\bibitem{Bertolami:2014wua}
Miller~Bertolami MM, Melendez BE, Althaus LG, Isern J.
\newblock \textit{JCAP} 1410:069 (2014)

\bibitem{Altarelli:2014ola}
Altarelli G.
\newblock \textit{Nucl. Instrum. Meth.} A742:56 (2014)

\bibitem{Feng:2013pwa}
Feng JL.
\newblock \textit{Ann. Rev. Nucl. Part. Sci.} 63:351 (2013)

\bibitem{Sikivie:2006ni}
Sikivie P.
\newblock \textit{Lect. Notes Phys.} 741:19 (2008)

\bibitem{Raffelt:2006cw}
Raffelt GG.
\newblock \textit{Lect. Notes Phys.} 741:51 (2008)

\bibitem{Isern:2008nt}
Isern J, Garcia-Berro E, Torres S, Catalan S.
\newblock \textit{Astrophys. J.} 682:L109 (2008)

\bibitem{Isern:2010wz}
Isern J, Garcia-Berro E, Althaus L, Corsico A.
\newblock \textit{Astron. Astrophys.} 512:A86 (2010)

\bibitem{Corsico:2012sh}
Corsico A, et~al.
\newblock \textit{JCAP} 1212:010 (2012)

\bibitem{Agashe:2014kda}
Olive K, et~al.
\newblock \textit{Chin. Phys.} C38:090001 (2014)

\bibitem{Sikivie:1983ip}
Sikivie P.
\newblock \textit{Phys. Rev. Lett.} 51:1415 (1983)

\bibitem{Sikivie:1985yu}
Sikivie P.
\newblock \textit{Phys. Rev.} D32:2988 (1985)

\bibitem{Dicke}
Dicke R.
\newblock \textit{Rev. Sci. Instr.} 17:268 (1946)

\bibitem{Sikivie:2010bq}
Sikivie P.
\newblock \textit{Phys. Lett.} B695:22 (2011)

\bibitem{vanBibber:2003sv}
van Bibber K, Kinion S.
\newblock \textit{Phil. Trans. Roy. Soc. Lond.} A361:2553 (2003)

\bibitem{DePanfilis:1987dk}
De~Panfilis S, et~al.
\newblock \textit{Phys. Rev. Lett.} 59:839 (1987)

\bibitem{Wuensch:1989sa}
Wuensch W, et~al.
\newblock \textit{Phys. Rev.} D40:3153 (1989)

\bibitem{Hagmann:1990tj}
Hagmann C, Sikivie P, Sullivan N, Tanner D.
\newblock \textit{Phys. Rev.} D42:1297 (1990)

\bibitem{Tada2006488}
Tada M, et~al.
\newblock \textit{Phys. Lett. A} 349:488  (2006)

\bibitem{Muck}
Muck M, et~al.
\newblock \textit{Applied Physics Letters} 72 (1998)

\bibitem{Muck:1999nra}
Muck M, et~al.
\newblock \textit{Appl. Phys. Lett.} 75:3545 (1999)

\bibitem{Muck2}
Muck M, Clarke J.
\newblock \textit{Journal of Applied Physics} 88 (2000)

\bibitem{Asztalos:2009yp}
Asztalos S, et~al.
\newblock \textit{Phys. Rev. Lett.} 104:041301 (2010)

\bibitem{Hoskins:2011iv}
Hoskins J, et~al.
\newblock \textit{Phys. Rev.} D84:121302 (2011)

\bibitem{Shokair:2014rna}
Shokair T, et~al.
\newblock \textit{Int. J. Mod. Phys.} A29:1443004 (2014)

\bibitem{CastellanosLehnert}
Castellanos-Beltran MA, Lehnert KW.
\newblock \textit{Applied Physics Letters} 91: (2007)

\bibitem{Castellanos}
Castellanos-Beltran MA, et~al.
\newblock \textit{Nat Phys} 4:929 (2008)

\bibitem{Xi}
Xi X, et~al.
\newblock \textit{Phys. Rev. Lett.} 105:257006 (2010)

\bibitem{Mallet}
Mallet F, et~al.
\newblock \textit{Phys. Rev. Lett.} 106:220502 (2011)

\bibitem{Lamoreaux:2013koa}
Lamoreaux S, van Bibber K, Lehnert K, Carosi G.
\newblock \textit{Phys. Rev.} D88:035020 (2013)

\bibitem{Schuster}
Schuster DI, et~al.
\newblock \textit{Nature} 445:515 (2007)

\bibitem{Wallraff}
Wallraff A, et~al.
\newblock \textit{Nature} 431:162 (2004)

\bibitem{Chen}
Chen YF, et~al.
\newblock \textit{Phys. Rev. Lett.} 107:217401 (2011)

\bibitem{Rybka:2014cya}
Rybka G, Wagner A  arXiv:1403.3121 [physics.ins-det] (2014)

\bibitem{Sikivie:2013laa}
Sikivie P, Sullivan N, Tanner D.
\newblock \textit{Phys. Rev. Lett.} 112:131301 (2014)

\bibitem{Baker:2011na}
Baker OK, et~al.
\newblock \textit{Phys. Rev.} D85:035018 (2012)

\bibitem{Horns:2012jf}
Horns D, et~al.
\newblock \textit{JCAP} 1304:016 (2013)

\bibitem{Budker:2013hfa}
Budker D, et~al.
\newblock \textit{Phys. Rev.} X4:021030 (2014)

\bibitem{Graham:2013gfa}
Graham PW, Rajendran S.
\newblock \textit{Phys. Rev.} D88:035023 (2013)

\bibitem{Linde:1987bx}
Linde AD.
\newblock \textit{Phys. Lett.} B201:437 (1988)

\bibitem{Wilczek:2012it}
Wilczek F  arXiv:1204.4683 [hep-th] (2012)

\bibitem{Arvanitaki:2014wva}
Arvanitaki A, Baryakhtar M, Huang X  arXiv:1411.2263 [hep-ph] (2014)

\bibitem{Arvanitaki:2010sy}
Arvanitaki A, Dubovsky S.
\newblock \textit{Phys. Rev.} D83:044026 (2011)

\bibitem{Arvanitaki:2009fg}
Arvanitaki A, et~al.
\newblock \textit{Phys. Rev.} D81:123530 (2010)

\bibitem{Graham:2011qk}
Graham PW, Rajendran S.
\newblock \textit{Phys. Rev.} D84:055013 (2011)

\bibitem{Stadnik:2013raa}
Stadnik Y, Flambaum V.
\newblock \textit{Phys. Rev.} D89:043522 (2014)

\bibitem{Stadnik:2014ala}
Stadnik YV, Flambaum VV  arXiv:1409.2986 [hep-ph] (2014)

\bibitem{Hong:1991fp}
Hong J, Kim JE.
\newblock \textit{Phys. Lett.} B265:197 (1991)

\bibitem{Roberts:2014cga}
Roberts B, et~al.
\newblock \textit{Phys. Rev.} D90:096005 (2014)

\bibitem{Roberts:2014dda}
Roberts B, et~al.
\newblock \textit{Phys. Rev. Lett.} 113:081601 (2014)

\bibitem{Moody:1984ba}
Moody J, Wilczek F.
\newblock \textit{Phys. Rev.} D30:130 (1984)

\bibitem{Vasilakis:2008yn}
Vasilakis G, Brown J, Kornack T, Romalis M.
\newblock \textit{Phys. Rev. Lett.} 103:261801 (2009)

\bibitem{Burghoff:2011zz}
Burghoff M, et~al.
\newblock \textit{J. Phys. Conf. Ser.} 295:012017 (2011)

\bibitem{Ledbetter:2012xd}
Ledbetter M, Romalis M, Jackson-Kimball D.
\newblock \textit{Phys. Rev. Lett.} 110:040402 (2013)

\bibitem{Heil:2013tpa}
Heil W, et~al.
\newblock \textit{Annalen Phys.} 525:539 (2013)

\bibitem{Tullney:2013wqa}
Tullney K, et~al.
\newblock \textit{Phys. Rev. Lett.} 111:100801 (2013)

\bibitem{Arvanitaki:2014dfa}
Arvanitaki A, Geraci AA.
\newblock \textit{Phys. Rev. Lett.} 113:161801 (2014)

\bibitem{Nelson:2011sf}
Nelson AE, Scholtz J.
\newblock \textit{Phys. Rev.} D84:103501 (2011)

\bibitem{Raffelt:1999tx}
Raffelt GG.
\newblock \textit{Ann. Rev. Nucl. Part. Sci.} 49:163 (1999)

\bibitem{Andriamonje:2007ew}
Andriamonje S, et~al.
\newblock \textit{JCAP} 0704:010 (2007)

\bibitem{Redondo:2013wwa}
Redondo J.
\newblock \textit{JCAP} 1312:008 (2013)

\bibitem{Buchmuller:1989rb}
Buchm\"uller W, Hoogeveen F.
\newblock \textit{Phys. Lett.} B237:278 (1990)

\bibitem{Paschos:1993yf}
Paschos EA, Zioutas K.
\newblock \textit{Phys. Lett.} B323:367 (1994)

\bibitem{Creswick:1997pg}
Creswick RJ, et~al.
\newblock \textit{Phys. Lett.} B427:235 (1998)

\bibitem{Avignone:1997th}
Avignone F.~T. I, et~al.
\newblock \textit{Phys. Rev. Lett.} 81:5068 (1998)

\bibitem{Morales:2001we}
Morales A, et~al.
\newblock \textit{Astropart. Phys.} 16:325 (2002)

\bibitem{Bernabei:2001ny}
Bernabei R, et~al.
\newblock \textit{Phys. Lett.} B515:6 (2001)

\bibitem{Ahmed:2009ht}
Ahmed Z, et~al.
\newblock \textit{Phys. Rev. Lett.} 103:141802 (2009)

\bibitem{Armengaud:2013rta}
Armengaud E, et~al.  arXiv:1307.1488 [astro-ph.CO] (2013)

\bibitem{Cebrian:1998mu}
Cebri{\'a}n S, et~al.
\newblock \textit{Astropart. Phys.} 10:397 (1999)

\bibitem{Avignone:2010zn}
Avignone III FT, Creswick RJ, Nussinov S  arXiv:1002.2718 [astro-ph.CO] (2010)

\bibitem{Ljubicic:2004gt}
Ljubicic A, Kekez D, Krecak Z, Ljubicic T.
\newblock \textit{Phys. Lett.} B599:143 (2004)

\bibitem{Derbin:2011gg}
Derbin A, et~al.
\newblock \textit{Phys. Rev.} D83:023505 (2011)

\bibitem{Derbin:2011zz}
Derbin A, Muratova V, Semenov D, Unzhakov E.
\newblock \textit{Phys. Atom. Nucl.} 74:596 (2011)

\bibitem{Derbin:2012yk}
Derbin A, Drachnev I, Kayunov A, Muratova V.
\newblock \textit{JETP Lett.} 95:379 (2012)

\bibitem{Bellini:2012kz}
Bellini G, et~al.
\newblock \textit{Phys. Rev.} D85:092003 (2012)

\bibitem{Moriyama:1995bz}
Moriyama S.
\newblock \textit{Phys. Rev. Lett.} 75:3222 (1995)

\bibitem{Krcmar:1998xn}
Krcmar M, et~al.
\newblock \textit{Phys. Lett.} B442:38 (1998)

\bibitem{Krcmar:2001si}
Krcmar M, et~al.
\newblock \textit{Phys. Rev.} D64:115016 (2001)

\bibitem{Derbin:2009jw}
Derbin A, et~al.
\newblock \textit{Phys. Lett.} B678:181 (2009)

\bibitem{Zioutas:2004hi}
Zioutas K, et~al.
\newblock \textit{Phys. Rev. Lett.} 94:121301 (2005)

\bibitem{vanBibber:1988ge}
van Bibber K, McIntyre PM, Morris DE, Raffelt GG.
\newblock \textit{Phys. Rev.} D39:2089 (1989)

\bibitem{Arik:2008mq}
Arik E, et~al.
\newblock \textit{JCAP} 0902:008 (2009)

\bibitem{Lazarus:1992ry}
Lazarus DM, et~al.
\newblock \textit{Phys. Rev. Lett.} 69:2333 (1992)

\bibitem{Inoue:2002qy}
Inoue Y, et~al.
\newblock \textit{Phys. Lett.} B536:18 (2002)

\bibitem{Moriyama:1998kd}
Moriyama S, et~al.
\newblock \textit{Phys. Lett.} B434:147 (1998)

\bibitem{Inoue:2008zp}
Inoue Y, et~al.
\newblock \textit{Phys. Lett.} B668:93 (2008)

\bibitem{Zioutas:1998cc}
Zioutas K, et~al.
\newblock \textit{Nucl. Instrum. Meth.} A425:480 (1999)

\bibitem{Kuster:2007ue}
Kuster M, et~al.
\newblock \textit{New J. Phys.} 9:169 (2007)

\bibitem{Abbon:2007ug}
Abbon P, et~al.
\newblock \textit{New J. Phys.} 9:170 (2007)

\bibitem{Aune:2013pna}
Aune S, et~al.
\newblock \textit{JINST} 9:P01001 (2014)

\bibitem{Aune:2013nza}
Aune S, et~al.
\newblock \textit{JINST} 8:C12042 (2013)

\bibitem{Aune:2011rx}
Arik E, et~al.
\newblock \textit{Phys. Rev. Lett.} 107:261302 (2011)

\bibitem{Arik:2013nya}
Arik M, et~al.
\newblock \textit{Phys. Rev. Lett.} 112:091302 (2014)

\bibitem{Andriamonje:2009dx}
Andriamonje S, et~al.
\newblock \textit{JCAP} 0912:002 (2009)

\bibitem{Andriamonje:2009ar}
Andriamonje S, et~al.  arXiv:0904.2103 [hep-ex] (2009)

\bibitem{Barth:2013sma}
Barth K, et~al.
\newblock \textit{JCAP} 1305:010 (2013)

\bibitem{Brax:2011wp}
Brax P, Lindner A, Zioutas K.
\newblock \textit{Phys. Rev.} D85:043014 (2012)

\bibitem{Baum:2014rka}
Baum S, et~al.  arXiv:1409.3852 [astro-ph.IM] (2014)

\bibitem{Irastorza:2011gs}
Irastorza IG, et~al.
\newblock \textit{JCAP} 1106:013 (2011)

\bibitem{Armengaud:2014gea}
Armengaud E, et~al.
\newblock \textit{JINST} 9:T05002 (2014)

\bibitem{Irastorza:1567109}
Irastorza IG.
\newblock {The International Axion Observatory IAXO. Letter of Intent to the
  CERN SPS committee}.
\newblock Tech. Rep. CERN-SPSC-2013-022. SPSC-I-242, CERN, Geneva (2013)

\bibitem{Shilon:2012te}
Shilon I, Dudarev A, Silva H, Kate H.
\newblock \textit{IEEE Trans. Appl. Supercond.} 23 (2012)

\bibitem{nustar2013}
Harrison FA, et~al.
\newblock \textit{Astrophysical Journal} 770:103 (2013)

\bibitem{doi:10.1117/12.2024476}
Jakobsen AC, Pivovaroff MJ, Christensen FE.
\newblock \textit{Proc. SPIE} 8861:886113 (2013), proc. SPIE 8861, Optics for
  EUV, X-Ray, and Gamma-Ray Astronomy VI, 886113 (2013)

\bibitem{Irastorza2011_EAS}
Irastorza I, et~al.
\newblock \textit{EAS Publications Series} 53:147 (2012)

\bibitem{Dafni:2012fi}
Dafni T, et~al.
\newblock \textit{J. Phys. Conf. Ser.} 375:022003 (2012)

\bibitem{Dafni:2012zz}
Dafni T, et~al.
\newblock \textit{J. Phys. Conf. Ser.} 347:012030 (2012)

\bibitem{trexwebpage}
{T-REX project web page: \url{http://gifna.unizar.es/trex/}}

\bibitem{Isern:2008fs}
Isern J, Catalan S, Garcia-Berro E, Torres S.
\newblock \textit{J.Phys.Conf.Ser.} 172:012005 (2009)

\bibitem{Corsico:2012ki}
Corsico AH, et~al.  arXiv:1205.6180 [astro-ph.SR] (2012)

\bibitem{redondo_patras_2014}
{J. Redondo, talk at Patras Workshop on Axions, WIMPs and WISPs, CERN, June
  2014. \url{http://axion-wimp2014.desy.de/}}

\bibitem{Ehret:2010mh}
Ehret K, et~al.
\newblock \textit{Phys. Lett.} B689:149 (2010)

\bibitem{Ehret:2009sq}
Ehret K, et~al.
\newblock \textit{Nucl. Instrum. Meth.} A612:83 (2009)

\bibitem{Purcell:1946}
Purcell E.
\newblock \textit{Phys. Rev.} 69:681 (1946)

\bibitem{Haroche:1989}
Haroche S, Kleppner D.
\newblock \textit{Phys.Today} 42:24 (1989)

\bibitem{Haroche:1991kj}
Haroche S.
\newblock \textit{Phys. World} 4N3:33 (1991)

\bibitem{Hoogeveen:1990vq}
Hoogeveen F, Ziegenhagen T.
\newblock \textit{Nucl. Phys.} B358:3 (1991)

\bibitem{Fukuda:1996kwa}
Fukuda Y, Kohmoto T, Nakajima Si, Kunitomo M.
\newblock \textit{Prog. Cryst. Growth Charact.Mater.} 33:363 (1996)

\bibitem{Mueller:2009wt}
Mueller G, Sikivie P, Tanner DB, van Bibber K.
\newblock \textit{Phys. Rev.} D80:072004 (2009)

\bibitem{Jaeckel:2007ch}
Jaeckel J, Ringwald A.
\newblock \textit{Phys. Lett.} B659:509 (2008)

\bibitem{Caspers:2009cj}
Caspers F, Jaeckel J, Ringwald A.
\newblock \textit{JINST} 4:P11013 (2009)

\bibitem{Betz:2013dza}
Betz M, et~al.
\newblock \textit{Phys. Rev.} D88:075014 (2013)

\bibitem{Ballou:2014myz}
Ballou R, et~al.  arXiv:1410.2566 [hep-ex] (2014)

\bibitem{Battesti:2010dm}
Battesti R, et~al.
\newblock \textit{Phys. Rev. Lett.} 105:250405 (2010)

\bibitem{Inada:2013tx}
Inada T, et~al.
\newblock \textit{Phys. Lett.} B722:301 (2013)

\bibitem{Bahre:2013ywa}
B{\"a}hre R, et~al.  arXiv:1302.5647 [physics.ins-det] (2013)

\bibitem{Black:2001}
Black E.
\newblock \textit{Am. J. of Phys.} 69:79 (2001)

\bibitem{Lita:2010}
Lita A.E. ea.
\newblock \textit{Proc. SPIE (Advanced Photon Counting Techniques IV)} 76810D
  (2010)

\bibitem{Miller:2011}
Miller A.J. ea.
\newblock \textit{Optics Express Vol. 19} 10:9102 (2011)

\bibitem{Dreyling-Eschweiler:2014eya}
Dreyling-Eschweiler J  DESY-THESIS:2014-016 (2014)

\bibitem{Dreyling-Eschweiler:2014mxa}
Dreyling-Eschweiler J  arXiv:1409.6992 [physics.ins-det] (2014)

\bibitem{Dreyling-Eschweiler:2014MO}
Dreyling-Eschweiler J. ea.
\newblock \textit{submitted to Journal of Modern Optics}  (2014)

\bibitem{Bottura:2012uka}
Bottura L, de~Rijk G, Rossi L, Todesco E.
\newblock \textit{IEEE Trans. Appl. Supercond.} 22:4002008 (2012)

\bibitem{Todesco:2014gna}
Todesco E, Bottura L, de~Rijk G, Rossi L.
\newblock \textit{IEEE Trans. Appl. Supercond.} 24:4004306 (2014)

\bibitem{Arias:2010bh}
Arias P, Jaeckel J, Redondo J, Ringwald A.
\newblock \textit{Phys. Rev. D} 82:115018 (2010)

\end{thebibliography}

\end{document}